\documentclass[manuscript]{acmart}

\AtBeginDocument{%
  \providecommand\BibTeX{{%
    \normalfont B\kern-0.5em{\scshape i\kern-0.25em b}\kern-0.8em\TeX}}}

    \usepackage{array}
\newcolumntype{P}[1]{>{\raggedright\let\newline\\\arraybackslash\hspace{0pt}}m{#1}}

\usepackage{float}
\usepackage{color}
\usepackage{booktabs}
\usepackage{textcomp}
\usepackage{dblfloatfix}
\usepackage{makecell}
\usepackage{url}
\usepackage{todonotes}
\usepackage{relsize}
\usepackage{subcaption}
\usepackage[bottom]{footmisc}

\newcommand*{\zx}{\textcolor{black}}
\newcommand*{\x}{\textcolor{black}}

\setcopyright{acmcopyright}
\copyrightyear{2020}
\acmYear{2020}
\acmDOI{10.1145/1122445.1122456}

\acmConference[Woodstock '18]{Woodstock '18: ACM Symposium on Neural
  Gaze Detection}{June 03--05, 2018}{Woodstock, NY}
\acmBooktitle{Woodstock '18: ACM Symposium on Neural Gaze Detection,
  June 03--05, 2018, Woodstock, NY}
\acmPrice{15.00}
\acmISBN{978-1-4503-XXXX-X/18/06}

\author{Ziang Xiao}
\email{ziang.xiao@jhu.edu}
\affiliation{%
  \institution{Johns Hopkins University}
  \city{Baltimore}
  \state{MD}
}
\affiliation{%
  \institution{Microsoft Research}
  \city{Montreal}
  \state{QC} 
}

\author{Q. Vera Liao}
\affiliation{%
  \institution{Microsoft Research}
  \city{Montreal}
  \state{QC} 
}
\email{veraliao@microsoft.com}

\author{Michelle X. Zhou}
\affiliation{%
  \institution{Juji. Inc.}
  \city{San Jose}
  \state{CA} 
}
\email{mzhou@acm.org}

\author{Tyrone Grandison\texorpdfstring{$^*$}{}}
\thanks{$^*$This research work was conducted as part of New Voices in Sciences, Engineering, and Medicine program of the US National Academies.}
\affiliation{%
  \institution{The Data-Driven Institute}
  \city{Seattle}
  \state{WA} 
}
\email{tgrandison@data-driven.institute}
\author{Yunyao Li\texorpdfstring{$^*$}{}}
\affiliation{%
  \institution{Apple}
  \city{San Jose}
  \state{CA}
}
\email{yunyaoli@apple.com}
\begin{document}

\title{Powering an AI Chatbot with Expert Sourcing to Support Credible Health Information Access}


\renewcommand{\shortauthors}{Xiao et al.}
\renewcommand{\shorttitle}{AI Chatbot for Credible Health Information}

\begin{abstract}
During a public health crisis like the COVID-19 pandemic, a credible and easy-to-access information portal is highly desirable. It helps with disease prevention, public health planning, and misinformation mitigation. However, creating such an information portal is challenging because 1) domain expertise is required to identify and curate credible and intelligible content, 2) the information needs to be updated promptly in response to the fast-changing environment, and 3) the information should be easily accessible by the general public; which is particularly difficult when most people do not have the domain expertise about the crisis. In this paper, we presented an expert-sourcing framework and created {\tt Jennifer}, an AI chatbot, which serves as a credible and easy-to-access information portal for individuals during the COVID-19 pandemic. {\tt Jennifer} was created by a team of over 150 scientists and health professionals around the world, deployed in the real world and answered thousands of user questions about COVID-19. We evaluated {\tt Jennifer} from two key stakeholders’ perspectives, expert volunteers and information seekers. We first interviewed experts who contributed to the collaborative creation of {\tt Jennifer} to learn about the challenges in the process and opportunities for future improvement. We then conducted an online experiment that examined {\tt Jennifer}’s effectiveness in supporting information seekers in locating COVID-19 information and gaining their trust. We share the key lessons learned and discuss design implications for building expert-sourced and AI-powered information portals, along with the risks and opportunities of misinformation mitigation and beyond.
\end{abstract}


\maketitle

\section{Introduction}
Public health crises, such as the COVID-19 pandemic, pose a major global threat to humankind. When such a crisis occurs, individuals actively seek information to make sense of the situation, mitigate the chronic uncertainty about the disease, and learn precautionary measures to protect themselves and others \cite{chao2020media,gunderson2021peer}. Although recent development in Information and Communication Technologies (ICT) accelerates information collection and dissemination during a crisis, it creates new challenges for information seekers, including information overload \cite{hagar2015crisis} and the prevalence of misinformation \cite{swar2017information,chae2016avoids}. Without proper guidance and support, individuals may turn to unreliable, even harmful, information that threatens themselves or society. In this study, we present an expert-sourcing framework to create AI chatbots that support the general public's information-seeking with credible and intelligible information.

A credible and easy-to-access information portal would help people make informed decisions and facilitate effective disease control, prevention, and public protection \cite{berland2001health,rochwerg2020misinformation,safieddine2016corporate,swire2020public}. It serves as an information collection and filtering agency and saves individuals time and effort to locate and identify credible information. It is especially critical at the early stage of a crisis when information is scarce and rapidly evolving. However, creating such an information portal is challenging. First, the complexity of a crisis requires a diverse set of domain expertise (e.g., epidemiology, public policy, psychology, etc.) to locate reliable information sources, verify the credibility of a new piece of information, and effectively communicate complex information to the general public \cite{arif2016information,boulos2011crowdsourcing}. However, such a team is difficult to recruit and domain experts are often occupied with many duties during a crisis. Second, the fast-changing scientific and public knowledge makes it important while expensive to keep the information up-to-date. A content curator needs to monitor information scattered across multiple resources (e.g., government websites). Additionally, the proliferated misinformation demands special attention and a fast reaction as information seekers rely on the portal to verify and debunk misinformation \cite{rochwerg2020misinformation}. The chronic uncertainty of a public health crisis makes this challenge long-lasting. Third, retrieving the needed information requires a significant amount of effort from the information seeker \cite{agree2015s}. During a crisis, especially at its early stage, information seekers may not have sufficient knowledge to organize effective search strategies. If the information portal can not provide affordances (e.g., search recommendations) to scaffold the search effort, information seekers may turn to information avoidance \cite{hagar2015crisis} or expose themselves to misinformation \cite{shklovski2010technology}. 

We present an expert-sourcing framework for building an AI chatbot for effective information triage. Our framework provides multiple benefits. First, the decentralized nature of an expert-sourcing framework encourages board participation and welcomes experts with a diverse skill set \cite{retelny2014expert}. We adopted a hierarchical structure \cite{retelny2017no} with a two-stage verification process to enable small team collaboration and assure information quality. Second, board participation enables fast iteration. The content can update quickly in response to the rapid-changing environment. Also, fast iteration makes the chatbot-building process dynamic. Our expert team could bootstrap daily conversation logs and improve the chatbot based on people's needs and feedback. Third, the interactivity of an AI chatbot facilitates information searching and drives engaging experiences \cite{rosset2020leading, tavakoli2020generating,bickmore2016improving,vtyurina2017exploring}. In a turn-by-turn chat, information seekers could build up complex information queries, clarify their information needs, and provide feedback to the chatbot to improve content quality. Compared to other common information portals such as social media or web search engines, a chatbot could provide a direct and concise answer without requiring information seekers to go through a long list of results \cite{miner2020chatbots}. Prior studies also showed that a chatbot interface can help engender trust \cite{sproull1996interface,folstad2017chatbots,altay2021information}, which is important when the information seeker is facing uncertainty during a public health crisis. Moreover, current technologies allow people to create and maintain a chatbot easily without extensive technical backgrounds \cite{jujidoc}, and embed it in many existing platforms, e.g., Facebook Messenger, and devices, e.g., Amazon Alexa, without extra development efforts.

During COVID-19, we applied our framework and, with the help of over 150 experts worldwide, created {\tt Jennifer}, an AI-powered chatbot that can answer people's COVID-19 questions. We deployed {\tt Jennifer} to the real world in early March 2020, the same month when COVID-19 was declared a pandemic by the World Health Organization. In a period of six months, {\tt Jennifer} answered more than 2000 questions in over 1200 chat sessions. Through this real-world deployment, we demonstrated the feasibility of directly sourcing the global scientific community's expertise for public benefit without the need for intermediaries, and to help improve public trust in science.

We evaluated our framework through the lens of our two major stakeholders, \textit{expert volunteers} who used our framework to build, deploy, and maintain {\tt Jennifer}; \textit{information seekers} who interacted with {\tt Jennifer} for their information needs. We asked two research questions, 

\begin{itemize}
    \item \textbf{RQ1:} How could we better support the creation and maintenance of an information portal, in the form of an AI chatbot, during a public health crisis?
\end{itemize}

We interviewed nine expert volunteers who contributed to the collaborative creation of {\tt Jennifer}. We summarized four major challenges and opportunities regarding the scalability and sustainability of our framework, including updating obsolete content update, effective health information communication, chatbot testing, and emotional support among volunteers.

\begin{itemize}
    \item \textbf{RQ2:}  How effective is {\tt Jennifer} in supporting people seeking COVID-19 information?
\end{itemize}

We conducted an online experiment with 77 participants and compared {\tt Jennifer} with a search engine, the most common way for people to get information on the internet, in COVID-19 information-seeking tasks. The results showed {\tt Jennifer} can better aid the information seeker's effort to locate credible information and gain their trust. 

Our contribution is three-fold. First, we proposed an expert-sourcing framework to create AI-powered chatbots as a credible and easy-to-access information portal in reaction to public health crises. We applied our framework and developed {\tt Jennifer} during COVID-19. With a team of over 150 expert volunteers, {\tt Jennifer} successfully answered over 2000 questions in a period of six months. Second, we dived into the creation process of {\tt Jennifer} through an interview study with expert volunteers. We distilled the challenges faced by our volunteers and gauged ideas for further improvement of our framework. Third, we evaluated {\tt Jennifer} and demonstrated its effectiveness in supporting people's information-seeking, driving higher user satisfaction, and gaining people's trust. We summarized design implications for an efficient process to create credible and easy-to-access information portals during a crisis, and discussed risks and opportunities of fighting misinformation with AI-powered chatbots and beyond.

\section{Related Work}
\subsection{Seeking Information during Crises}
In the event of a public health crisis, people seek information to make sense of the situation to inform their decision-making, reduce anxiety caused by the long-lasting uncertainty, and learn precautionary measures to protect themselves and others \cite{chao2020media}. Information and Communication Technologies (ICT) now accelerate information propagation during a crisis. People can access millions of information through social networks and get their questions answered by online articles or strangers on online forums. However, information abundance may create information overload that hinders effective information-seeking \cite{hagar2015crisis}. First, information overload reduces the information seeker's cognitive capacity to identify misinformation and creates stress, fatigue, and negative emotions \cite{guo2020information}. The cognitive capacity is further limited during a crisis when information seekers face many uncertainties and a rapidly changing environment. Second, information seekers need more time and effort to locate reliable sources and retrieve the answer to their questions. Shklovski et al. \cite{shklovski2010technology} found that individuals tend to find back channels and expose themselves to misinformation if they cannot access credible information on time. Information portals support information seekers by curating data from various sources \cite{choi2015real, kogan2018conversations} and filtering high-quality information \cite{alam2018processing,li2019identifying,kaufhold2020mitigating}.

The recent development of natural language interfaces has allowed people to seek information with a computer through conversations \cite{bickmore2016improving,rosset2020leading}. In a conversational search, people type questions in natural language, and the computer responds with complete sentences \cite{radlinski2017theoretical}. Studies have shown various benefits of a conversational search; including higher search efficiency and better user engagement. The natural language interface encourages information exchange. A conversational agent can process an individual's question on the fly and ask for clarification if the query is unclear \cite{rosset2020leading, tavakoli2020generating}. Similarly, individuals could ask follow-up questions if they are not satisfied with the system's answer. Conversational search reduces people's burden to find the right search terms that are normally used in conventional Web search \cite{bickmore2016improving,vtyurina2017exploring}. Trippas et al. \cite{trippas2018informing} showed that verbal communications encourage users to actively seek more specific information using complex queries. Moreover, conversational agents like chatbots can provide personified experiences. Their anthropomorphic features could help attract user attention and gain user trust \cite{sproull1996interface,folstad2017chatbots}. In this study, we create an AI-powered chatbot as the information portal to aggregate and filter credible information from large volumes of noise.

\subsection{Powering Crowdsourcing with Experts}
Crowd-sourcing decomposes complex tasks into simple pieces and calls a big crowd online for contributions. The power of the crowd has created high-quality datasets for machine learning models \cite{yuen2011survey}, scaled technical systems \cite{kanhere2013participatory}, and augmented system functionalities \cite{bernstein2010soylent}. The citizenry is a powerful force that enables ICT to play a transformational role during a crisis \cite{palen2010vision, poblet2018crowdsourcing}. Starbird et al. demonstrated the effectiveness of sourcing the crowds to identify first-hand information tweets from people who are local to a mass disruption event \cite{starbird2012learning}. Ludwig et al. situated crowd teams during crises through public displays \cite{ludwig2017situated}. However, it is challenging for a crowd-sourced team to complete complex tasks that require domain-specific skills or expertise \cite{retelny2017no,gong2019social}. One solution is sourcing experts with adequate domain knowledge, e.g., expert-sourcing. Expert-sourcing enables a wider range of tasks that require domain expertise, including prototype design, course building, film animation, and software development \cite{retelny2014expert,chen2016towards}. However, given the scarcity of experts, an effective and scalable expert-sourcing framework requires carefully designed workflow and infrastructure support \cite{white2015expertise}. In this study, we demonstrated the feasibility of sourcing experts during a public health crisis and shared key design implications toward a more effective expert-sourcing framework. 

\subsection{Building Chatbots that can Answer People's Questions}
Many chatbot platforms have been built to facilitate the creation of chatbots that can answer people's questions. In general, chatbot platforms can be divided into two types \cite{xiao2020if}. The rule-based system uses pre-defined rules to capture the user's exact questions and deliver a pre-defined answer \cite{chatfuel}. Although rule-based platforms provide a reliable way to answer people's questions, little natural language understanding means that every question needs to be pre-defined, which makes them costly to scale \cite{xiao2020if}. The AI-based systems provide more machine learning (ML) capabilities to capture user questions and generate answers but require high-quality training data and ML expertise \cite{dialogflow}. While large language models, like GPT-3, allow people to build a capable AI chatbot without extensive ML resources, its pre-trained knowledge base and hallucination tendency limit its ability to answer people's questions about a novel crisis. To solve the scalability problem in rule-based systems and the data scarcity problem in AI-based systems, researchers also explored crowd-powered chatbots \cite{lasecki2013chorus}. Crowd workers could directly power a rule-based chatbot by writing responses or selecting the most appropriate pre-defined responses \cite{lasecki2013chorus,huang2018evorus}. However, when building a chatbot for people's questions about a novel public health crisis, domain expertise is required to ensure response quality. In our framework, we extend a hybrid approach that supports both rules and ML models with an expert sourcing framework to build a chatbot that can answer COVID-19 questions with credible and intelligible information.

\section{System Overview}
\subsection{Design Considerations: Chatbot as the Information Portal}
Information overload \cite{hagar2015crisis}, the prevalence of misinformation \cite{swar2017information,chae2016avoids}, and ineffective information retrieval \cite{miller2012online} impede an individual's information-seeking during a public health crisis. \zx{To address the above challenges, we built a chatbot as an information portal with the following design considerations:}

\subsubsection{Ease of Access} \x{When a crisis happens, an information seeker should easily satisfy their needs without facing information overload or conflicting messages. Meanwhile, with misinformation spreading predominately on social media and the Web, the public must have an accessible information source to fact-check. Therefore, the information portal should provide information to the general public in an easily accessible manner across different platforms (e.g., Web and social media). Easy access is also beneficial for the cascading dissemination of accurate information. Compared to a stand-alone website, a chatbot could be embedded into any existing website or social media. Additionally, the conversational interface of a chatbot provides multiple benefits for information seekers, including search efficiency and personalized experiences \cite{rosset2020leading, tavakoli2020generating,bickmore2016improving,vtyurina2017exploring}. Through a natural language conversation, information seekers can formulate queries in their own language. And a chatbot could help information seekers to build up a complex query turn-by-turn. Such a personalized experience could reduce information-seeking fatigue and make the information easy to access.}

\subsubsection{Rapid Development} During a public health crisis, the information portal has to be built in a short time to best inform the general public and win the race against fast-spreading misinformation. Compared to a website, which often requires full-stack development work, let alone the search component, today's chatbot building platforms allow a chatbot builder to take advantage of AI and deploy a chatbot that can answer people's questions quickly without extensive technical backgrounds. Moreover, a chatbot could be embedded into different platforms and devices, e.g., social media, voice assistants, and websites, with minimal effort.

\subsubsection{Quality Assurance} The information portal has to provide credible and authentic information. Any small piece of misinformation that comes from the portal will misguide our users and undermine the platform's reputation and ultimately lead to its end. The portal must be built with and safeguarded by a rigorous process that ensures the platform information’s quality, authenticity, and scientific validity. With quality assurance, a centralized information portal will reduce the information seeker's cognitive load during the information retrieval process. In addition, the information seeker's feedback is crucial to consider. Compared to other forms of interaction, a turn-by-turn chat enables information seekers could naturally give explicit feedback about information quality.

\subsubsection{Communication Effectiveness} The information portal has to deliver public health information in a clear and intelligible way since public health information often contains domain-specific medical knowledge. Without proper communication, the general public won't act upon it, which will negatively impact disease control and prevention. It is important for the portal to ensure information is communicated in a simple, natural, consumable, and empathetic manner. A chatbot could naturally deliver complex medical information in plain language and allow information seekers to ask for clarification. Moreover, compared to search engines or social media where the information seeker needs to face a long list of results, the chatbot could provide a more concise and direct response which reduces potential information overload and aids communication effectiveness \cite{miner2020chatbots}. 

\subsubsection{Ease of Update and Extension} In reaction to the fast-changing situation during a novel public health crisis, the portal's knowledge base and system functions (e.g., multilingual support) should be update-able and extensible with minimal effort. Ideally, the system should be managed and maintained by volunteers without a deep technical background. In our framework, we choose chatbot-building platforms with a graphical interface so the chatbot can be updated and maintained easily without technical backgrounds.
 
\subsection{System Architecture}
\begin{figure*}[t]
    \centering
    \includegraphics[width=0.80\textwidth]{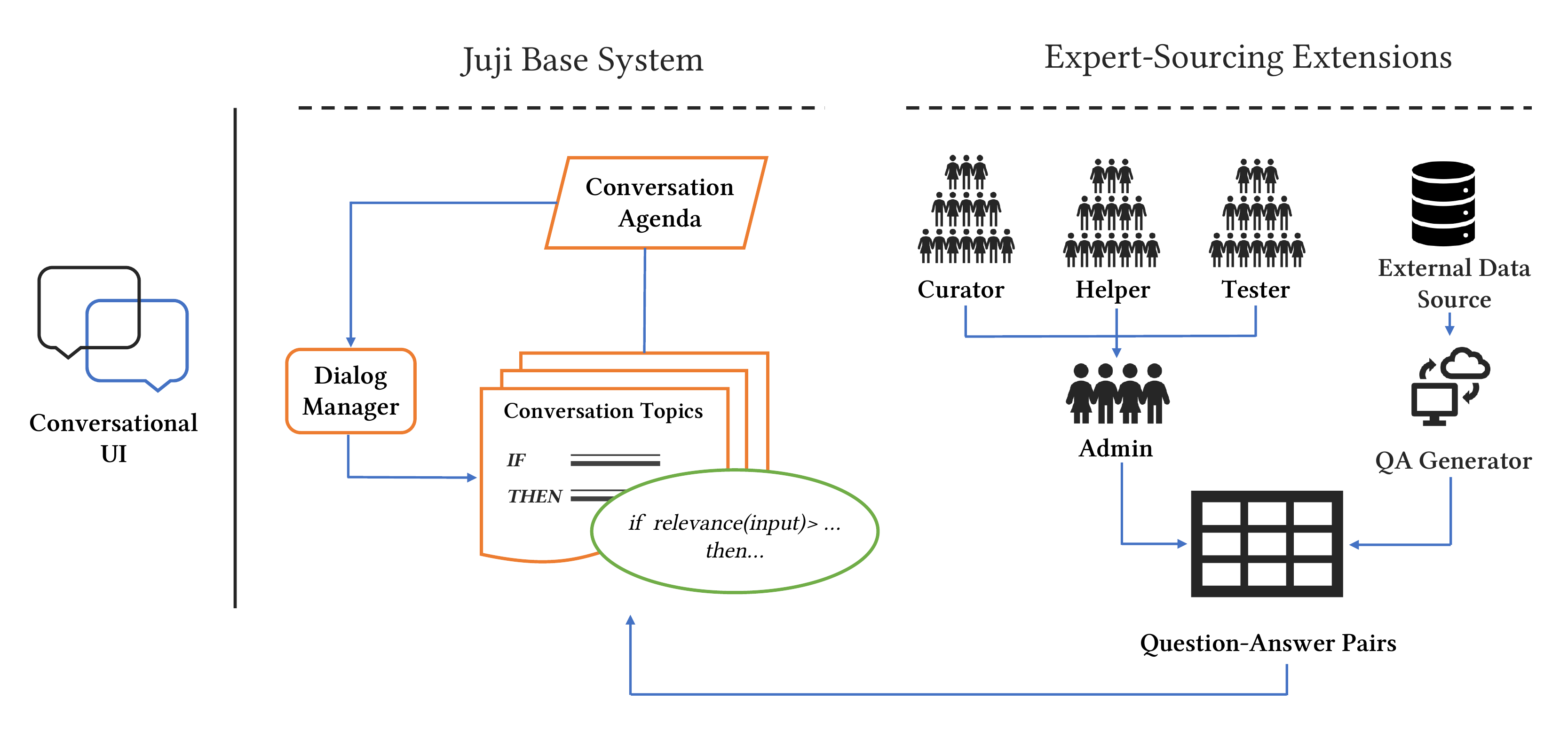}
    \vspace{-0.05in}
    \caption{The figures show the architecture overview of {\tt Jennifer}. Our expert sourcing framework extends Juji's base system's ability to answer people's questions about COVID-19. Through a hierarchical structure, expert volunteers were divided into four groups, Admins, Curator, Helper, and Tester, and worked together to create QA pairs with credible and easy-to-access information. Those QA pairs extend Juji's dialogues system's ability to answer people COVID-19 related questions.}
    \label{fig:overview}
\end{figure*}

\zx{During the global COVID-19 pandemic, we applied our framework and developed {\tt Jennifer} \cite{li-etal-2020-jennifer}. Fig \ref{fig:overview} depicts the overall architecture of {\tt Jennifer}. {\tt Jennifer} builds on the Juji base system for dialog management \cite{zhou2019trusting}. We chose Juji for its ability to handle the mixed-initiative conversation, to support both rules and machine learning models, and to be extended with minimal effort. The Juji platform supports no-code-AI. The team could build an AI-powered chatbot that can understand people's questions and retrieve answers without extensive AI expertise, which benefits later maintenance and updates. The Juji platform generates URLs for both Web and Facebook deployment, which enables \textit{public and easy access to {\tt Jennifer}}. Given a user question, Juji uses a pre-trained machine learning model to identify relevant questions with known answers and returns an answer or a follow-up question; depending on its confidence level (See more, Sec. \ref{ChatDesign}). The main capabilities of {\tt Jennifer} come from the Question-Answer (QA) pairs generated by the extensions specifically implemented for {\tt Jennifer} with two modes of ingestion:}

\begin{itemize}
    \item \textbf{Expert-sourced}: This mode relies on a repository of Frequently Asked Questions gathered from reliable sources such as the Centers for Disease Control and Prevention (CDC) \footnote{https://www.cdc.gov/}, the World Health Organization (WHO) \footnote{https://www.who.int/}, the University of Washington Bothell \footnote{https://www.washington.edu/coronavirus/}, and the Federation of American Scientists \footnote{https://fas.org/}. The questions are provided by the users and volunteers of {\tt Jennifer}, many based on the FAQs. The answers are manually curated by the volunteers of {\tt Jennifer} via a rigorous process (See \ref{workflow}).
    
    \item \textbf{Automated generation}: Often, users of {\tt Jennifer} ask questions on specific statistics such as the number of confirmed cases in a state, city, or country. Since the answers to these questions are constantly changing, it is labor-intensive to curate answers manually. Instead, we built a QA Generator to automatically create such QA pairs, based on structured data pulled from reliable sources, such as the CDC, daily and populate question templates derived from the expert-sourced questions.
\end{itemize}

\subsection{Sourcing Experts and Professionals over the World}
Sourcing expertise around the globe allows us to build the information portal in a decentralized, fast, and reliable manner. Given the unknown and complex nature of a public health crisis, we leveraged expert-sourcing \cite{retelny2014expert} to curate content (e.g., QA pairs) and operate the portal. We made open calls on social media and networks within scientific communities to recruit volunteers who are medical experts, scientists, engineers, technologists, or specialists. The expert-sourcing framework aids the rapid development of the portal and enables a team with diverse domain expertise. 

\subsubsection{Operational Structure}
To ensure quality, reproducibility, and efficiency, we crafted processes and defined roles wrapped into an operational structure that enables the efficient delineation of tasks and preserves scientific integrity. 

\textit{\textbf{Admins}} recruit new team members, coordinate all roles, check unanswered questions, and manage available tasks. Admins validate question-answer pairs, first for scientific validity and then for language fluency and naturalness. To increase the quality of the resulting QA pairs, the same admin cannot perform both the validity and fluency checks for the same QA pair. 

\textit{\textbf{Curators}} manage QA pairs. They take new, unanswered questions, research current answers from reputable and trustworthy sources, and craft intelligible answers with supporting evidence. Curators also update obsolete answers. Given the novelty of any public health crisis, this is a critical task to identify credible answers. We only assign this role to a small number of volunteers who have verified expertise (e.g., health professionals, biologists, and virologists).

\textit{\textbf{Helpers}} verify answers, make notes if it needs further investigation, and revise answers for readability. They also take existing questions and generate many possible question formulations, i.e., alternative questions. This step provides data to train ML models to better recognize people’s questions and to enhance {\tt Jennifer}'s ability to deliver the best corresponding answers. We open this role to a broader set of volunteers who have technology backgrounds.

\textit{\textbf{Testers}} test the portal by trying to retrieve the newly added questions with variations to ensure updates were properly implemented. They also evaluate answers for freshness, accuracy, and readability. They also monitor the system for other possible quality issues, e.g., format issues. We invite a broader set of volunteers, especially those who have experience with creative writing and chatbot evaluation.

\subsubsection{Workflow}
\label{workflow}
First, Admins and Curators leverage the trusted information sources to seed the system with its initial set of QA pairs.  Once the chatbot is deployed, the Admins periodically export the unanswered questions from the system. Admins create a task for curators when a new, unanswered question comes in. Curators can sign up for a task and start to curate an answer. Once the research on a question and its answer have been done, Helpers are flagged to check the answers created by Curators for credibility and quality. If any problem emerges, Helpers make a note, send a flag, and update Curators. Helpers also create questions that are semantically similar to the question that they are helping with; to enhance {\tt Jennifer}'s natural language understanding capability. Helpers flag when their QA pairs are ready for testing. Admins upload those flagged entries into the pre-deployment platform. Testers use the pre-deployment platform interfaces to chat with {\tt Jennifer} and ensure that there are no further issues with the QA pairs to be deployed. Admins perform the final check on all QA pairs that pass testing and mark them for deployment, which is executed periodically on all QA pairs that have successfully gone through the workflow. It should be noted that everyone in the team can signal to Admins and Curators if anything in the question bank is outdated. 

\subsubsection{Answer Quality Assurance and Communication Effectiveness}

To be included in the system, each answer needs to satisfy the following criteria:

\begin{itemize}
    \item \textit{Easy to understand}: The information is presented in language intelligible by the general public.
    \item \textit{Accuracy and Openness}: The answers must be backed up by evidence from reliable sources, including references or links to such sources, and be verified by at least one trusted volunteer medical expert. Furthermore, scientific understanding of a public health crisis is quickly evolving; it is important to be explicit about potential uncertainty in the answers, e.g., presenting evidence from both sides of view.
    \item \textit{Demonstration of Empathy}: The language provided in the answers should emulate natural empathetic conversation, and must acknowledge factors, such as stress or anxiety, experienced by the users to help foster trust.
\end{itemize}

\begin{figure}[t]
\begin{subfigure}{0.29\textwidth}
    \includegraphics[width=0.9\linewidth]{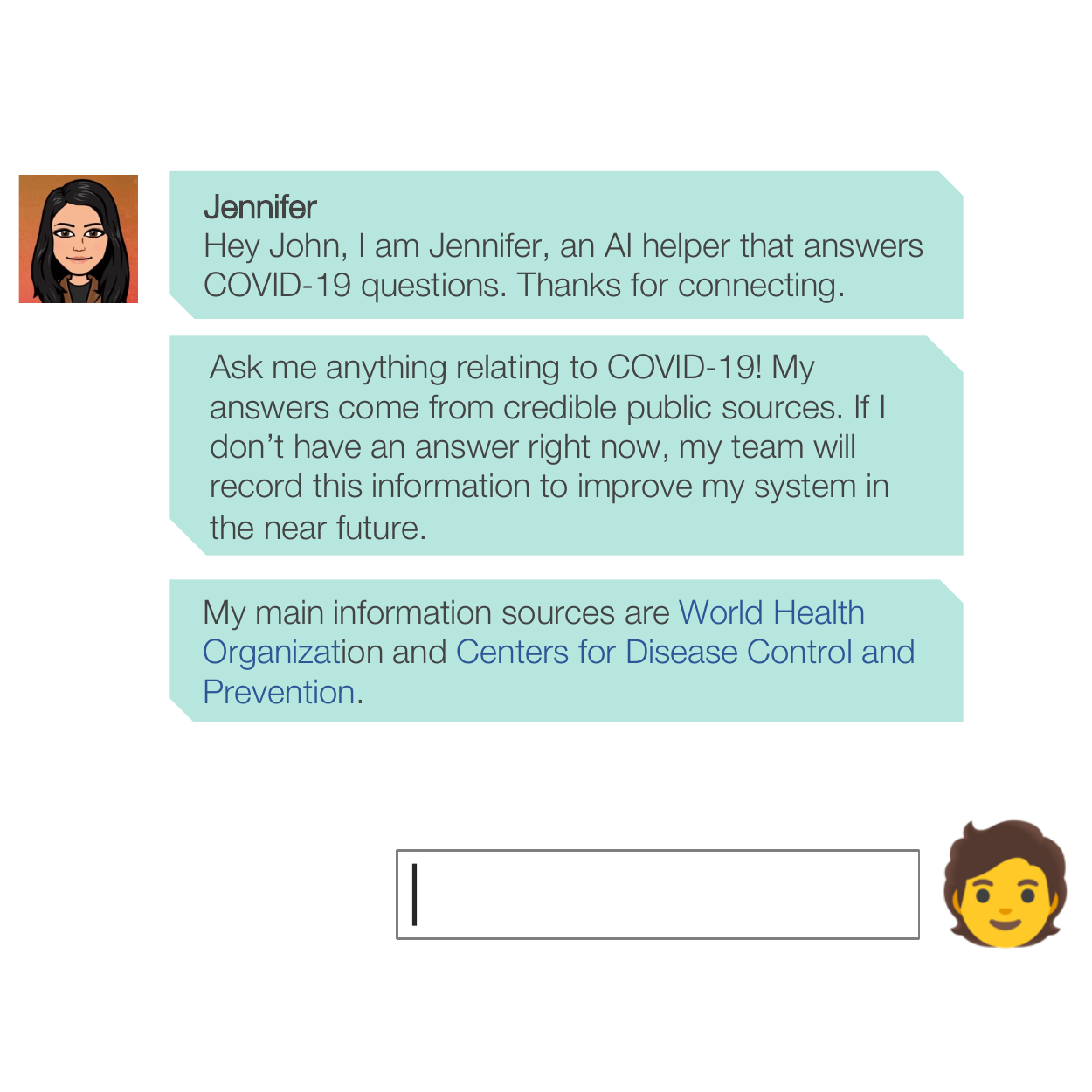}
    \label{fig:welcome}
    \caption{Jennifer opens the conversation with a self-introduction.}
\end{subfigure}
\begin{subfigure}{0.29\textwidth}
    \includegraphics[width=0.9\linewidth]{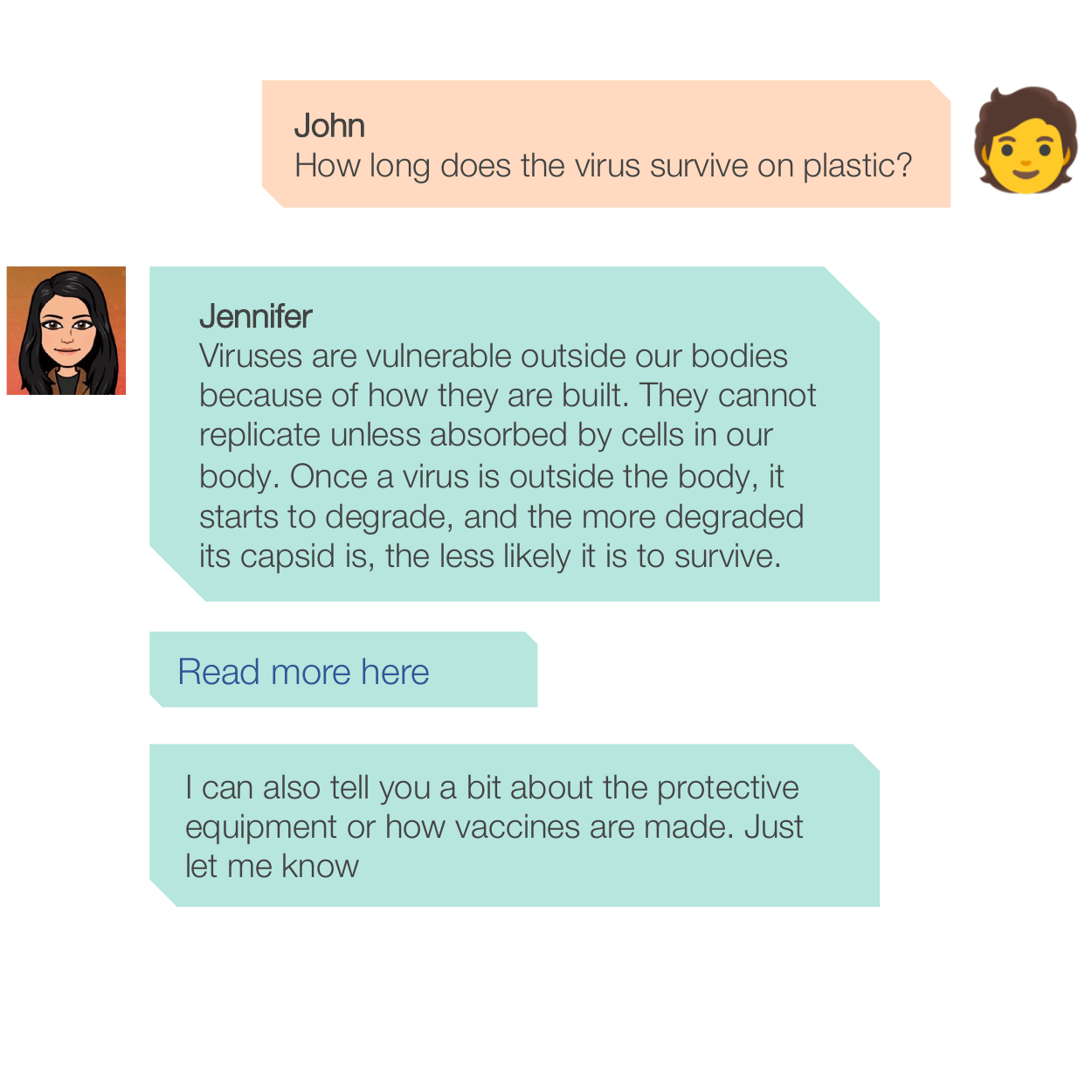}
    \caption{Jennifer recommends relevant questions}
    \label{fig:relevant}
\end{subfigure}
\begin{subfigure}{0.29\textwidth}
    \includegraphics[width=0.9\linewidth]{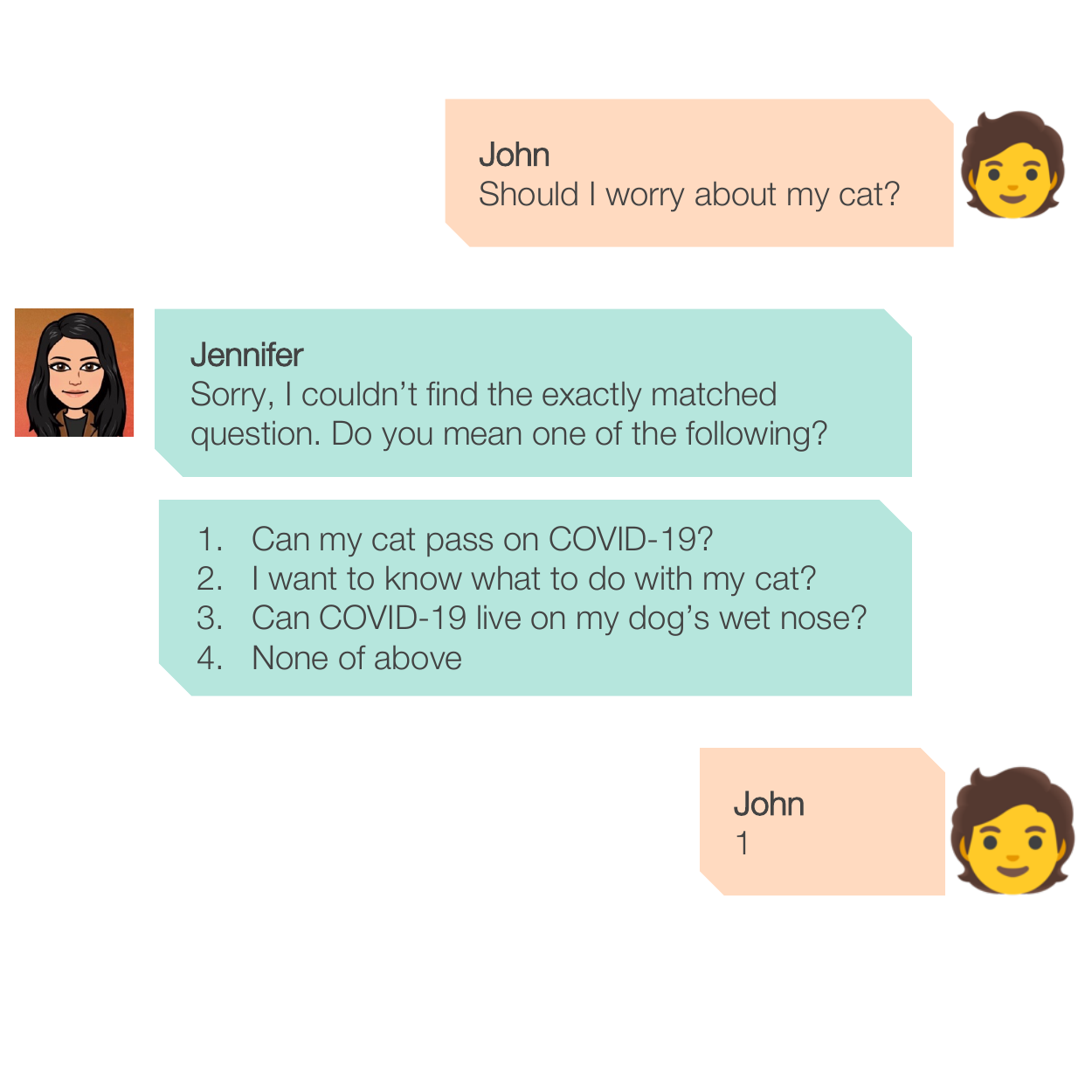}
    \label{fig:clarification}
    \caption{Jennifer clarifies user's questions.}

\end{subfigure}

\caption{The figures demonstrate three examples of how {\tt Jennifer} interacts with information seekers.}
\label{fig:design}
\end{figure}

\subsection{Chat Design}
\label{ChatDesign}
\subsubsection{Supporting Two-way adaptation} When {\tt Jennifer} was first launched in early March, most people knew little about COVID-19 or its impact. Thus, {\tt Jennifer} started with a ``menu'' to inform users about its existing knowledge on the most important topics. After answering a question, {\tt Jennifer} also volunteers information on additional topics that it knows. This design aims to address two challenges: 1) the user may not know how to get started or lack the knowledge to ask additional questions; 2) {\tt Jennifer} will never be perfect; there will always be questions that it cannot answer. By informing users about what it knows, users are more likely to ask questions that {\tt Jennifer} can answer. If {\tt Jennifer} is unsure about how to answer a question, it will recommend similar questions to give users a chance to obtain desired answers as well as learn more about {\tt Jennifer}’s capabilities. Fig. \ref{fig:relevant} shows how it expresses its uncertainty regarding the user’s question but proceeds to recommend a list of relevant inquiries. {\tt Jennifer} will improve its response to similar questions based on user interactions. 

\subsubsection{Fostering mixed-initiative interaction} {\tt Jennifer} aims at fostering mixed-initiative interactions. On the one hand, it proactively solicits questions from users. On the other hand, it allows users to initiate their questions at any time during the chat flow. Such mixed-initiative interactions keep users engaged while enabling users to obtain information at their own pace.

\subsection{Real World Deployment}
At the beginning of COVID-19, we quickly mobilized, created, and deployed {\tt Jennifer} to aid people's information-seeking efforts during the crisis. We sourced global scientific communities by sending open calls through our personal social networks, newsletters, and online articles. We deployed our {\tt Jennifer} on March 8th, 2020, which is around the same time that the World Health Organization officially announced COVID-19 as a global pandemic. For 6 months, from Mar 2020 to Aug 2020, we received 170 responses and recruited 159 scientists and professionals from 141 institutions worldwide. Most of them were engaged in the first three months, and then the involvement waned down as time goes. At its peak, we had a total of 181 volunteers working together \footnote{This included volunteers, members of the Juji team, and members of the initial cohort of the New Voices in Science, Engineering, and Medicine program from the National Academies.}.  By the end of August, the expert team had created over 10,000 QA pairs. During the course of the 6-month real-world deployment on both Facebook Messenger and New Voices Website \footnote{https://www.newvoicesnasem.org/}, {\tt Jennifer} handled 1,252 chat sessions. {\tt Jennifer} was asked 2,982 questions with an answer rate of 76 \%. On average, people interact with Jennifer for 3.83 minutes. During each session, people were asked an average of 2.38 questions. 

The real-world deployment demonstrates the feasibility of our framework in building AI-powered chatbots as information portals and aiding the general public's information-seeking effort. It also provides a valuable opportunity for us to further evaluate our framework through a holistic view. We ran two separate studies, which will be described later, about {\tt Jennifer} from both the system builder's perspective and the information seeker's perspective, two key stakeholders of our framework.

\section{Study 1: Diving into Jennifer's creation process.}

To answer our first research question, this study aimed to understand how the expert team used our framework to create and maintain {\tt Jennifer}. We contacted and interviewed nine expert volunteers (denoted as Ex\textbf{\#}) who contributed to the creation and maintenance process of {\tt Jennifer}. Out of those nine volunteers, five of them are women, and four of them are men. They served different roles: two Admins, three Curators, three Helpers, and two Testers. They are either scientists in computer science, biochemistry, molecular biology, and chemistry or professionals from the health and chatbot industry. The interview was conducted via Zoom online. Each interview lasted about 45 minutes. At the beginning of each interview, we asked our participants about their roles and tasks in building {\tt Jennifer}. We are particularly interested in what their workflow was and what tools were used. We then focused on the challenges they have experienced or been aware of. We also tried to elicit how they approached those challenges and what they wanted to improve. In the end, we asked for tools they would like to use to overcome the challenges. All interviews were transcribed and then analyzed through inductive coding and clustering to identify common challenges and opportunities for system support. Our analysis reveals the following four common themes mentioned by our participants. 

\subsection{Updating Obsolete Content}
The design goal of our framework is to make the most updated information accessible. However, information evolves rapidly during a public health crisis, especially pandemics like COVID-19, which we know little about. \textit{``So the biggest challenge for me was the ongoing information that was changing all the time, because we were learning new information about this virus.''} [Ex2].

Although in our framework, we have designed our knowledge base to be easily updated, curators have to keep following the content they have created to identify obsolete knowledge. When the volume grows, it becomes more challenging to keep everything updated. \textit{``we definitely have to go back to the question all the time, because you remember what question you was being asked to answer. So when something new come up, ..., I think I put the other one for the answer. So you'd come back. ... you just need to update it and put it back in the system.'' } [Ex1]. Although expert volunteers could notify each other when new information comes out, it is nearly impossible to keep track of every piece of information.\textit{``  I'm looking up information on that all the time. ...it is impossible [to keep up with everything]. '' } [Ex1].

During the interview, expert team members shared a few ideas for automation that may lift the burden off their shoulders. Overall, there are three categories of information that often need to update, statistics (e.g., \# of new cases), public policy (e.g., mask mandates, testing requirements, lockdown), and disease knowledge (e.g., safety protocol, symptoms, incubation period). Statistics needs most frequent updates, \textit{``We had to update, because this was when, in the beginning of the pandemic, ... , the numbers were constantly changing and people care.'' } [Ex8], but can be automated easily \textit{``You could automate the regular stats, you could automate, like how many daily new cases they are, x, y country and x, y county'' } [Ex5]. \textit{``Finding a reliable source is most important and [build] API ...'' } [Ex6]. Later in the process, for statistics-related questions, we applied a simple template-based approach to automatically generate the corresponding QA pairs with data pulled from CDC and WHO to minimize manual efforts. 

Obsolete public policy may have huge implications. \textit{``I always read news to check government websites to update those policies. I don't want people to get into trouble because of our answers'' } [Ex8]. Unlike statistics, no single API could offer updated public policies. One Helper shared an idea for monitoring and notification. \textit{``It will be helpful if something could, you know, like a security camera, look at those government websites and send notifications if something changed.'' } [Ex8]. 

Compared to the other two categories, knowledge about the disease is the most difficult to keep track of. First, no central outlet tracks new scientific publications. \textit{``we need to constantly look for new publications and they are everywhere.'' }[Ex3]. Second, it is difficult to determine if a piece of knowledge is obsolete. \textit{``like airborne. New evidence came out every day and still doesn't have a clear answer'' }[Ex1]. To keep knowledge about the disease updated, our expert volunteers want a summarizer for scientific papers that summarize, \textit{``if we can see all available evidence, it can make our decisions [to update a piece of knowledge] much easier.''}, and track the latest updates, \textit{``it could be based on citation.'' }[Ex8].

\subsection{Communicating Health Information Effectively}
Communicating health information to the general public is hard \cite{rowan1991simple}. Making the information easy to consume not only guides disease prevention but also makes people feel mentally reassured. \textit{``People come to Jennifer be, like my mother or my father, who doesn't have a scientific background or anything like that. They just worry. And they want to know. So I won't bombard them with jargon or scientific terms, I want to do is giving them the information that they can understand and be more assured. ''}[Ex1].

Although our framework has multiple safeguards (Curator, Helper, Tester, and Admin) to ensure the final answer is direct, concise, and easy to understand, team members found making the scientific language intelligible is not easy. One Curator mentioned \textit{``... It's tough for me to write answers that everyone could understand. I spent a lot of time to make sure the answer I wrote can be read easily, although I knew other people will edit my answer later''}[Ex6].    

Additionally, for a novel disease, not all questions can be answered at the moment with a definite answer. Expert team members found it challenging to communicate the information uncertainty and evolving knowledge. \textit{``Because sometimes we may know a little bit about something, but not definitely. So we didn't want to make a statement, because the data was still not 100\%. So the challenge was to find a way to communicate effectively without giving false information. ''}[Ex2]. The task is harder as the answers created by experts also need to fit {\tt Jennifer}'s conversational style.\textit{``Jennifer, to me, is someone very nice, very knowledgeable, very calm. It is not always easy to write the answer consistently in her tone.''}[Ex1].

In response to those challenges, experts voiced the need for a writing support tool. They envision such a tool to have several features, 1) providing real-time feedback or suggestions to make an answer easy to consume. \textit{``I would like to have a robot telling me how to make those things [health information] easy to understand.''}[Ex8]. The real-time feedback provided by the writing assistant could not only improves readability but also teach team members scientific writing skills. 2) automatically generating easy-to-understand answers. One Helper mentioned the potential of natural language generation models to translate medical jargon into languages that the general public could understand. \textit{``You know things like GPT-3. Maybe it can be used to generate answerers that easy to read. ''}[Ex6]. Although the automated methods may significantly reduce people's workload, I6 also raised concerns of imperfect AI and reflected the need for human oversight. \textit{``It may not be perfect and could be dangerous sometimes. We definitely need to look at it before putting it into the system. ''}[Ex6]. and 3) matching a chatbot's conversational style. \textit{``It will be great if some assistants could make the language style coherent.''}[Ex1]. It is challenging for a chatbot to have a coherent language style with hundreds of writers behind it. An automated language style checker would lift the burden on the writing style and help the expert team focus more on the information quality.

\subsection{Auditing and Testing the Chatbot}
Although an AI-powered chatbot could better understand users' questions, drive engaging experiences, and deliver rich conversations, the ``black-box'' nature of machine learning models behind an AI-powered chatbot posed the challenge of testing for our expert teams. All Testers mentioned the challenge of testing {\tt Jennifer} with updated content before deploying to the general public. Answering people's questions during a public health crisis where each individual's decision may have huge implications on disease prevention and self-protection leaves no room for mistakes. When a new QA pair is added or updated, to ensure quality, Testers need to test, first, if the new answer could be retrieved by the corresponding question or semantic similar questions, and second, if the new QA pair would interfere with previously implemented QA pairs. \textit{``I will ask the same question in two to maybe four or five different ways ... we ask for formulate questions in different types of ways. We use different terms or phrases. ... see if the bot picks up the answer.''}[Ex9]. A lot of repeated human labor is required to test comprehensively especially when there are over 10,000 QA pairs. \textit{``the difficulty is testing all paths, it's, it's very difficult to test our paths, because, ultimately, and I suppose this is why they put it out to a larger audience. ''}[Ex7]. One Tester mentioned the challenge of tracking testing progress. \textit{``I mean, mentally, I knew what I had asked. But sometimes I asked the same question more than once.''}[Ex9].

Also, testers test if the chatbot would work for different screens as {\tt Jennifer} could be easily accessed with many devices. \textit{``I tested from various devices, and various browsers to see if it was, it would function better or worse than others. I tested the responses to see if they stayed on screen with the amount of screen size that was provided. ''}[Ex7]

Our participants envision a testing and auditing tool to ensure quality. The most straightforward assistance is simply tracking the test progress. \textit{``It's cool if it can highlight what has been tested.''}[Ex9]. A more advanced tool can also run automatic test cases like the ones used in software testing. For example, each QA pair would be considered as a test case and the tool can audit the chatbot by enumerating all available test cases when an update is available. \textit{``the testing ... for instance, made some updates, the system can check first and leave uncertain ones for manual test.''}[Ex4]

\subsection{Building Emotional Support among Team Members}
The last emerging theme is the need for emotional support among team members. Unlike most crowdsourcing frameworks where workers mostly work individually without socially interacting with others, our expert team members expressed the need for emotional support from other team members and its implication for volunteer engagement. One of the Admins noted, \textit{``We need to keep the volunteers engaged, and they feel they are part of a bigger thing and other people also care about what they care about.''}[Ex4].

However, connecting team members, especially in a global team during a pandemic, is not easy. \textit{People are spread out in different countries, different time zones on have never met before. ..... I actually will try to have a brief chat, on zoom, or in Skype with everybody on to explain the project the scope on and make suggestions, or we could collaborate to each other to establish some point of relationship on that is, in my experience, important that people who collaborate also feel that they have a relationship together.} [Ex3].

Others also expressed the need to interact socially with other team members. \textit{``I think it would be more personal and more connected''}[Ex6]. A global public health crisis puts everyone in a stressful or even isolated environment. Working together as a team could positively impact an individual's mental health. \textit{`` I like the idea of being a part of something, like being involved because I'm totally kind of isolated from other people [during COVID-19]''}[Ex2]. 

To support interpersonal interactions, experts shared the idea of using teleconferencing tools to connect the global team \textit{``if we can, like, okay, let's have a Zoom meeting every month ''}[Ex6], or a shared virtual common space that everyone in the team could access, \textit{``something like a lobby would be wonderful, virtually of course''}[Ex8]. A virtual common space could satisfy various needs across team members, including building a personal connection, \textit{``I am curious who else are on the team''}[Ex6], providing mental support \textit{``it may make me feel less isolated''}[Ex2], sharing cheerful stories \textit{`` If we could do better, I will say maybe we should have a share more information to the volunteers about the impact of their work, right? ''}[Ex4], or encouraging scientific collaboration \textit{``It is a great chance to chat with other scientists with shared interests. New ideas may pop up.''}[Ex1].

\section{Study 2: Evaluating Jennifer in supporting people's information seeking.}
To answer our second research question on the effectiveness of {\tt Jennifer}, we turned our focus to information seekers who interact with {\tt Jennifer} for their information-seeking endeavor. We designed an online experiment and compared {\tt Jennifer} against a Web search engine in a COVID-19 information-seeking task. 

\subsection{Method}
We choose the Web search engine as our comparison platform for three reasons. First, to the best of our knowledge, although several chatbots have been built to answer people's questions about COVID-19 \cite{miner2020chatbots}, no prior study has evaluated such a chatbot like {\tt Jennifer} on its effectiveness in helping people access information during a public health crisis. Hence, no baseline chatbot we can use to compare {\tt Jennifer} with. Second, at the time of our study, existing chatbots for COVID-19 F\&Q either only support button-based interaction or are keyword-based that have limited capability to understand natural language. And the anthropomorphic characteristics are also limited. Hence, we chose the best keyword-based information retrieval method, a web search engine, as our baseline. Third, a Web search engine is one of the most popular ways of getting public health information online \cite{fahy2014quality}. When people have questions about COVID-19, people go to a Web search engine for an answer \cite{rochwerg2020misinformation}. In this study, we allow participants to choose a Web search engine based on their own preferences. 

\subsubsection{Information Seeking Task}
\label{sec:task}
The goal of the information-seeking task is to simulate a scenario where people need credible and intelligible COVID-19 information. We designed an information-seeking task to ask participants to find answers to five multiple-choice questions about COVID-19. To simulate the need for correct answers in a timely manner, we timed the task for 10 minutes and provided extra rewards if all questions were answered correctly. 

Each information-seeking task randomly drew five multiple-choice questions from a COVID-19 question bank. We curated the question bank based on three criteria. First, questions should have the proper difficulty level that demands that the participant seek external information (e.g., Whether wearing a mask could slow the spread won't be picked). Second, the question should ask about COVID-19 information that has meaningful implications on the general public's behavior for disease prevention and control (e.g., questions about Remdesivir's mechanism \footnote{https://en.wikipedia.org/wiki/Remdesivir} won't be picked). Third, the question should tap into COVID-19 information that is often targeted in misinformation campaigns where finding the correct answer requires extra efforts to combat misleading ones.

To select questions with the proper difficulty level, we first pulled 80 COVID-19-related questions from online sources such as The Guardian \footnote{https://www.theguardian.com/world/2020/apr/28/quiz-how-much-do-you-know-about-the-coronavirus}, Nebraska Medicine \footnote{https://www.nebraskamed.com/COVID/fact-check-part-1-covid-19-myths-and-misinformation-quiz} and Bloomberg \footnote{https://www.bloomberg.com/features/2020-coronavirus-quiz/}. Then, to learn the difficulty level of those questions, we compiled all questions together and sent them to 54 individuals through Amazon Mechanical Turk. We instructed participants to answer those questions with their own knowledge, and we stated that their performance wouldn't affect their reward. Additionally, we asked the participant to select the ``I don't know'' option if they were unsure about the answer. We calculated the accuracy for each question and used the first quartile ($M_{\text{accuracy}}=0.63$, $SD_{\text{accuracy}}=0.23$, $Q_{\text{1st}}=0.31$) as a cut off to select 20 questions that participants answered wrong or were unsure about. We then selected 18 out of the 20 questions as our final question set based on the second and third criteria listed above. The average accuracy for selected questions is $0.20$ (SD = $0.04$). To ensure the validity of our evaluation, the question selection process is independent of {\tt Jennifer}'s knowledge base.

The final question set tapped into four types of common coronavirus misconception or misinformation suggested by \cite{brennen2020types}, including virus transmission, virus origin, community spread, and public authority actions. The correct answers are backed by credible and reliable sources such as CDC \footnote{https://www.cdc.gov/} or WHO \footnote{https://www.who.int/}. Before deploying the task, we also performed our own search to make sure our answers reflected the best available information. We purposely removed questions related to vaccines due to the fast-evolving vaccine development during the study period.  

\subsection{Study Procedure}
We designed a within-subject study where the participants completed two information-seeking tasks with help from {\tt Jennifer} and a Web search engine. Participants interact with two tasks in random order with counterbalance. Our study has three sections. Upon consent, in the first section, participants completed a questionnaire about their attitudes toward chatbots and COVID-19.

In the second section, participants completed two COVID-19 information-seeking tasks (Described in \ref{sec:task}). Before the start of each task, we asked participants to open {\tt Jennifer} or the Web search engine in a separate browser window. We then instructed the participant to copy the first sentence said by {\tt Jennifer} or the first search result of the term ``COVID-19'' to a text box to make sure the chatbot or the search engine is ready for use. The pasted content was later used as an attention checker. During the task, participants could use {\tt Jennifer} or the web search engine to search for the correct answer. After each of the information-seeking tasks, participants were asked to complete a questionnaire about their experience and leave comments about the study. In the search engine task, we additionally asked what search engine was used.

We collected participants' demographic information in the third section. In the end, we debriefed our participants with the study purpose and correct answers to the information-seeking task from credible sources. 

\subsubsection{Measures}
We measured the effectiveness of {\tt Jennifer} in supporting information seekers to gather information about COVID-19 from four perspectives, 1) How accurate a participant's answers are, 2) How people trust the gathered information. 3) Time and effort spent to gather the information, and 4) How intelligible the gathered information is. All measures are on 5-point Likert scales from ``Strongly Disagree'' to ``Strongly Agree'', except answer accuracy.

\textit{Answer Accuracy:} The ability to help people find the correct answer is a key measure for information-seeking support. It is especially important during a public health crisis because inaccurate information may put people at risk. In each of the information-seeking tasks, the participants need to find answers to five COVID-19 questions using either {\tt Jennifer} or the web searching engine. We calculated the percentage of correct answers as answer accuracy ranging from 0 to 1.

\textit{Trust Intention:} Whether people trust the information provided by the portal suggests how the retrieved information would be consumed and future portal usage. If the portal retrieves accurate information, fostering such trust will help people resist contradicted misinformation, disseminate credible information, and act based on credible information \cite{bode2018see}. To measure the trust intent, we adapted an existing 7-item scale on trust intent to fit our needs \cite{mcknight2002impact}.

\textit{Time and Effort:} To aid information-seeking, the portal needs to retrieve information effortlessly. An effective information portal supports information seekers by providing effective and engaging search experiences. To measure the perceived time and effort, we adapted the ASQ scale with two items \cite{lewis1991psychometric} on time and effort, which were later combined into a single score by averaging.

\textit{Comprehensibility:} Comprehensibility is important when communicating medical information. The retrieved information wouldn't be useful if information seekers were not able to understand it. Communicating public health information in an intelligible way reduces information overload and misinterpretation. We asked our participants to rate on a 5-point scale if the results were intelligible.

\subsubsection{Attitudes and Demographics}
\textit{Prior attitudes towards chatbots:} People's prior experience with and attitude towards technology may affect their interaction with the technology \cite{nass1994computers}. In our study, we adapted questions from the Technology Acceptance Model \cite{van1997simple} to chatbot usage, which measures people’s attitudes towards a system from the perspective of Usefulness (5 items) and Satisfaction (4 items). We added questions to measure people’s pre-existing trust toward chatbots and their interaction frequency with conversational agents.

\textit{Prior attitudes towards COVID-19:} Existing literature indicates that people's prior attitude toward COVID-19 correlates with their information-seeking behavior and vulnerability towards misinformation \cite{czeisler2020public}. We used an existing scale \cite{czeisler2020public} to collect participants’ attitudes toward COVID-19 prevention policies prior to the study and their awareness of the prevalence of COVID-19 0misinformation. Both attitude questionnaires were asked in the first section of the study.  

\textit{Basic Demographics:} We collected basic demographic information, including age, gender, education level, and annual household income. 

\subsubsection{Recruitment}
We recruited our participants from Amazon Mechanical Turk in early 2021. We sent out our tasks in five batches over the course of two weeks: three on weekdays and two on the weekends to recruit a larger variety of participants. Participants were paid \$ 12.5/hr regardless of their performance. On average, they spent approximately 20 minutes on the task (M = 21.71 mins, SD = 8.32 mins). Our task is limited to English speakers and people who have a 95\% Approval Rate. Repeated answers were removed.

\subsubsection{Analysis Plan}
We analyzed our data using linear mixed-effect models for their robustness in modeling repeated measures for within-subject study designs. Since each participant in our study evaluates two information-seeking tools on the same set of measures, we treated each participant in the linear mixed effect model as a random effect \cite{faraway2016extending}. We treated \textit{Answer Accuracy}, \textit{Trust Intention}, \textit{Time and Effort}, and \textit{Comprehensibility} as dependent variables. The independent variable is whether the participant used {\tt Jennifer} or the search engine to perform the information-seeking task. We included participant's basic demographic, prior attitude and experience with chatbots, and prior attitude to COVID-19 as covariates to control for potential confounding effects as suggested by existing literature \cite{nass1994computers,czeisler2020public}.

\subsubsection{Limitations} We acknowledge the following limitations in our study design. \zx{Although search engine is one of the most prominent ways for people to gather information online during a crisis \cite{bento2020evidence,rovetta2020covid}, people also seek information online via other sources, like social media \cite{pennycook2020fighting,al2020lies}. On social media, information seekers may face different challenges as malicious users may deliberately spread misinformation and even disinformation through their interaction with the information seeker. Our study examines the effectiveness of {\tt Jennifer} when the information seeker is actively searching and distilling information online without direct interaction with other users. Even though today's search engines index user-generated content from social media like Twitter and Reddit, information seekers face different challenges on social media. Since {\tt Jennifer} could be easily integrated into social media sites such as Facebook messager, in the future, we ought to examine the role of {\tt Jennifer} in supporting information seekers on social media.}

The study was conducted in early 2021. Although the COVID-19 pandemic was still ongoing and the confirmed cases in the US were climbing, the general public had better knowledge about COVID-19 compared to the early stage. Many credible information portals have been created that grant people access to reliable information about COVID-19. The Web search engine can direct people to those reputable information portals. Additionally, compared to the early stage of COVID-19, more measures (e.g., highlighting information from credible sources) have been developed on web search engines to protect people from misinformation. Although we carefully designed our information-seeking task and selected questions with the consideration of difficulty level, the implication to real-world behavior, and vulnerability to misinformation, further studies are needed to understand how the information portal created by our framework would work at the very early stage of a public health crisis.

Our framework bootstrap chat sessions to progressively add QA pairs to {\tt Jennifer}. With such a design, although the chatbot could be built within a day and keep learning new content along the way, {\tt Jennifer}'s capability was limited at the beginning. By the time of our study, {\tt Jennifer} has been engaged in a 6-month-long active development, and our expert team has curated over 10,000 QA pairs. While, in general, people love {\tt Jennifer}, it is yet unknown how people would react to {\tt Jennifer} when her knowledge base is relatively scarce. Future studies are needed to understand how people would react to {\tt Jennifer} at different stages. 

Prior studies suggest that people's level of perceived anthropomorphism may affect their trust in the conversational agent \cite{go2019humanizing,xiao2021let}. In our study, although {\tt Jennifer}'s anthropomorphic features were the same for all participants, people's perceptions might be different. Different levels of perceived anthropomorphism in our study may influence people's trust in {\tt Jennifer}'s answers to their questions which may subsequently affect the effectiveness of {\tt Jennifer} in supporting the participant's information-seeking task. Since the current study design did not measure participants' perceived anthropomorphism of {\tt Jennifer}, it is important for future studies to parse out the effects of perceived anthropomorphism to better design an effective informational chatbot during crises.

\subsection{Results}
Overall, {\tt Jennifer} effectively supports people's information-seeking endeavor. Our results suggested that people would be able to find more correct answers with help from {\tt Jennifer} compared to the Web-search engine. People also reported a higher level of trust towards {\tt Jennifer}, {\tt Jennifer} is easy to use, and the information retrieved is intelligible.

\subsubsection{Participant Overview}
Out of the 90 participants we recruited, 77 (Denoted as P\textbf{\#}) completed the study and passed our attention and duplication check. Among those 77 participants, 30 identified as women, and 47 identified as men. The median education level was a Bachelor's degree. The median household income was between \$50,000 - \$ 100,000. And the median age of participants was between 25 - 34 years old. 59.74 \% (N = 46) of the participants indicated that they interacted with chatbots at least once per week. 3.90 \% (N = 3) of the participants said they had no recent interaction with a chatbot. Our participants considered chatbots generally Useful (M = 3.76, SD = 0.57), Satisfying (M = 3.98, SD = 0.90), and Trustworthy (M = 3.83, SD = 1.15). In general, participants supported COVID-19 prevention and control policies (M = 4.27, SD = 0.76) and were aware of the spreading of COVID-19 misinformation (M = 4.06, SD = 0.91). On average, our participants spent 7.32 mins (SD = 2.46) finishing the task with the Web-search engine, and 6.70 mins (SD = 1.81) with {\tt Jennifer}. Three participants in the Web-search engine and two participants in {\tt Jennifer} condition ran out of time. All participants answered at least four questions. In the task with the Web-search engine, the majority (97.40\%; N = 75) of the participants chose Google. One participant decided to use Bing, and one chose Yahoo. 

\subsubsection{People found more accurate answers with the help from {\tt Jennifer}}
An effective information portal should satisfy information seekers' queries. Our participants were able to find the correct answers with help from {\tt Jennifer} ($M_{accuracy}$ = 0.69; $SD_{accuracy}$ = 0.20). Given the task difficulty ($M_{accuracy}$ = 0.20; $SD_{accuracy}$ = 0.04; examined in the pilot study), {\tt Jennifer} could effectively aid information seeker's need for COVID-19 information.

\begin{figure}[t]

\begin{subfigure}{0.4\textwidth}
    \includegraphics[width=0.9\linewidth]{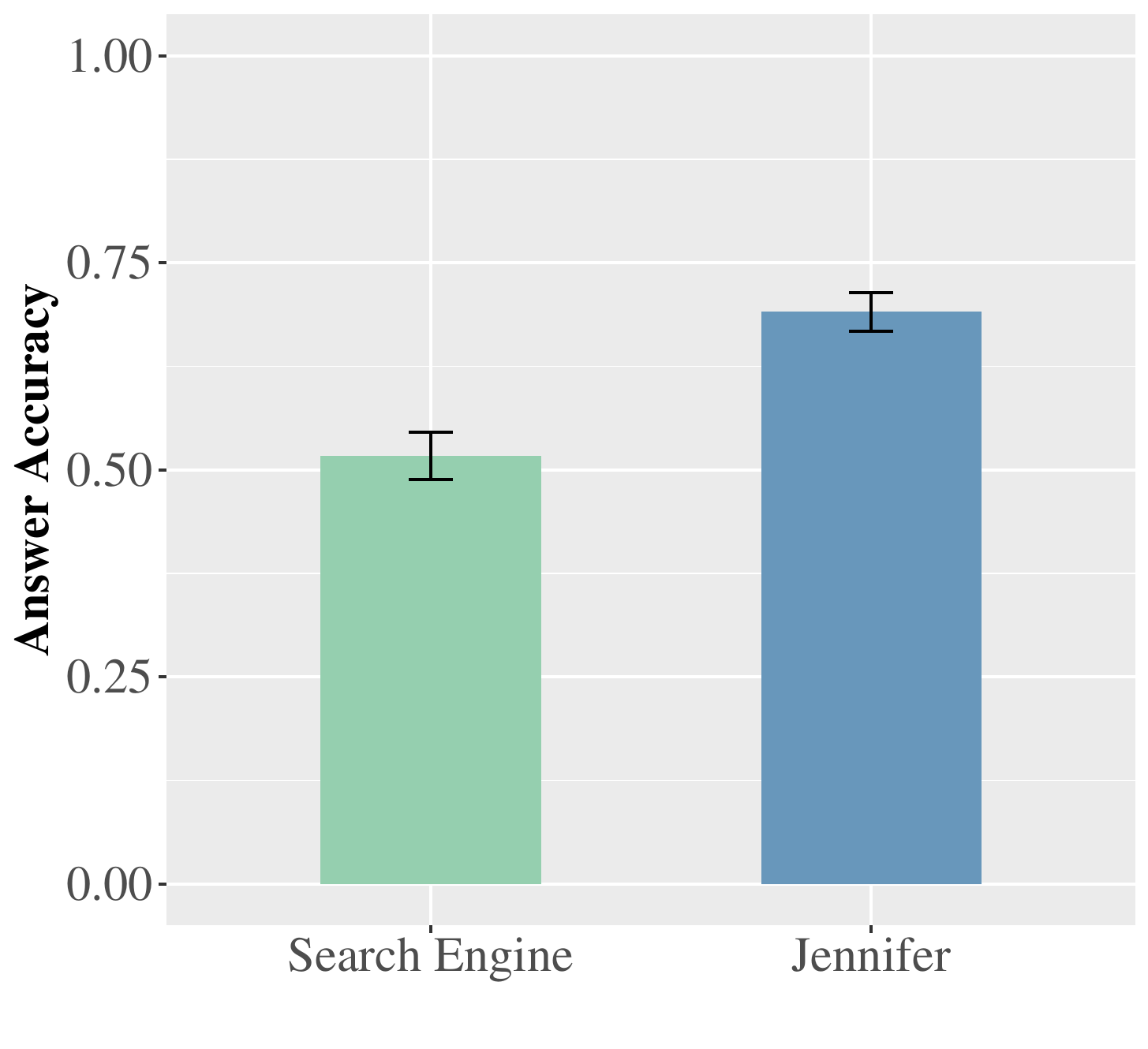}
    \caption{Participants’ answer accuracy}
    \label{fig:accuracy}
\end{subfigure}
\begin{subfigure}{0.4\textwidth}
    \includegraphics[width=0.9\linewidth]{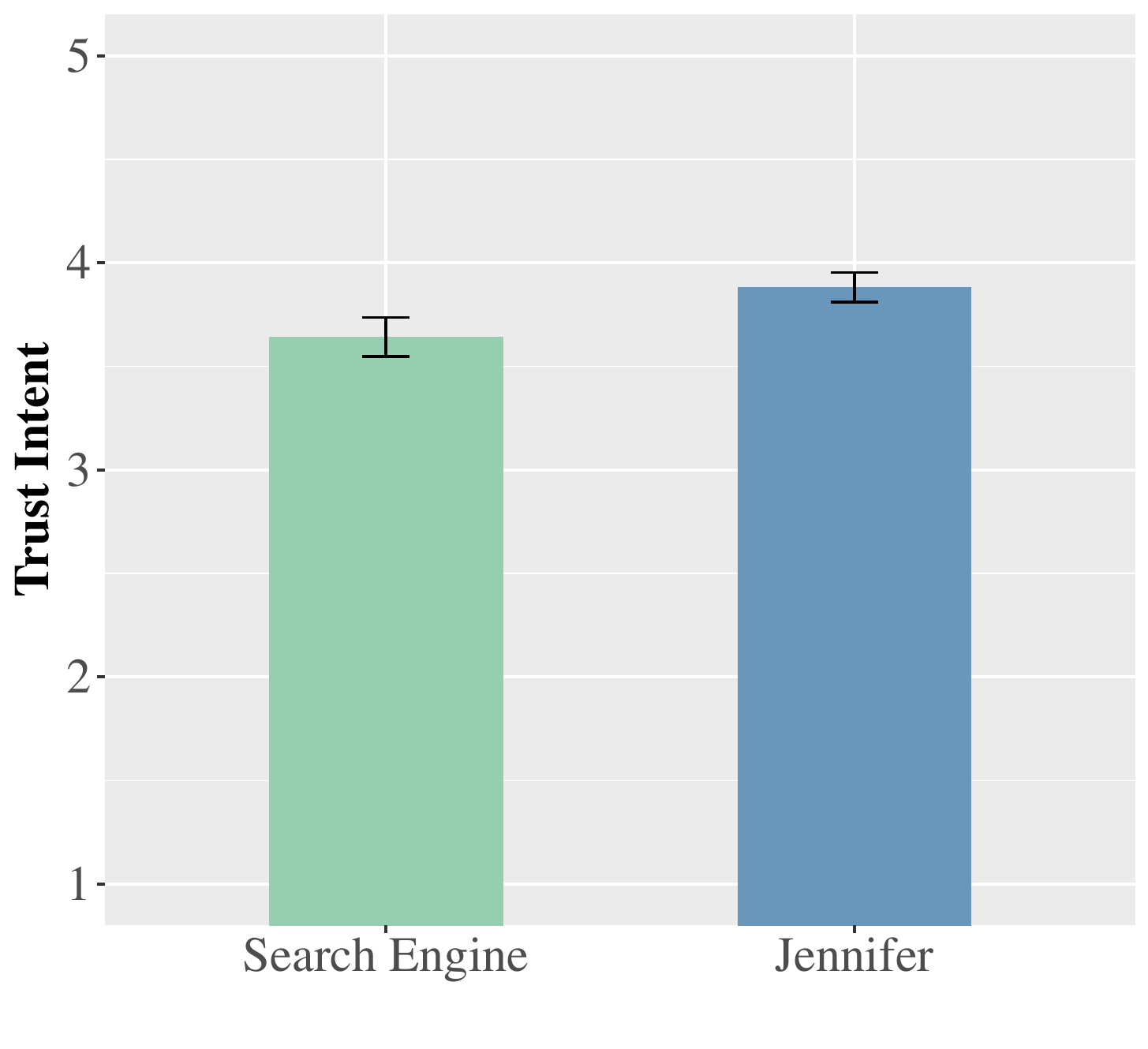}
    \label{fig:trust}
    \caption{Participants’ trust intent}
\end{subfigure}

\caption{The figures show contrasts between two conditions in terms of participants' answer accuracy and their trust intent. Main Findings: Compared to the Web-search engine, {\tt Jennifer} could better help our participants to locate correct answers. And our participants trust the information provided by {\tt Jennifer} more compared to what they found with a search engine.}
\label{fig:result1}
\end{figure}

We also found a significant main effect of the use of {\tt Jennifer} on the participant's information-seeking task performance (Fig~\ref{fig:result1}). Compared to using the Web-search engine ($M_{accuracy}$ = 0.52; $SD_{accuracy}$ = 0.25), by chatting with {\tt Jennifer} ($M_{accuracy}$ = 0.69; $SD_{accuracy}$ = 0.20), the participants were able to find the correct answers for more questions ($\beta$ = 0.18, SE = 0.03, t = 6.04, p < 0.01***). The effect is medium with a Cohen's d of 0.53. The results indicate the {\tt Jennifer} could better support information seekers to find COVID-19 information accurately compared to the Web-search engine. One of the potential reasons is that the expert-sourcing framework enables {\tt Jennifer} to aggregate information from multiple credible sources and deliver answers in an intelligible way. One participant commented, \textit{``I thought the answers were well thought out and explained well.''}[P1].

Compared to the Web-search engine, {\tt Jennifer} allows the users to phrase their questions in natural language. Consistent with prior research, people like to use the natural language interface to search for what they need \cite{rosset2020leading, tavakoli2020generating}. One participant noted, \textit{``The chatbot responded quickly to my questions. I liked how I was able to directly get answers to my questions by just typing them in.'' }[P49]. Some participants complained that it is sometimes difficult to find the right term to search and that search results from the Web-search engine were sometimes overwhelming, which makes them difficult to navigate and find the answer. For example, one participant commented, \textit{``I thought it was still a little challenging.  It was hard to find the exact phrase to find the information for a few of the questions and I didn't want to search too long. '' }[P27]. However, the natural language understanding capability of an AI chatbot is far from perfect. 15 participants mentioned in their comments that the chatbot won't be able to understand their questions fully and that they need to rephrase the questions or give up on that question. 

We found a significant effect of people's awareness of the prevalence of COVID-19 misinformation on their answers ($\beta$ = 0.38, SE = 0.11, t = 3.58, p < 0.01***) where people who are more aware of the prevalence of COVID-19 misinformation found more correct answers. Those participants may have a better pre-existing knowledge about COVID-19. The results also aligned with the previous evidence, which shows that people who are aware of misinformation might be better at navigating through misinformation and locating more credible information \cite{czeisler2020public}. No interaction effect was found between the use of {\tt Jennifer} and participants' awareness of COVID-19 misinformation.

\subsubsection{People trusted {\tt Jennifer} more}
Gaining people's trust is also crucial for effective information seeking. With trustworthy information, information seekers could save time and effort for cross-validation. Overall, our participants trusted the {\tt Jennifer} (M = 3.52; SD = 1.12) when gathering COVID-19 information. Being a trustworthy information portal, {\tt Jennifer} is able to effectively deliver credible and reliable COVID-19 information curated by expert volunteers. 

Additionally, the results on the trust intent scale showed a significant difference between the {\tt Jennifer} and the Web-search engine. When using the {\tt Jennifer}, people reported a significantly higher level of trust intent ($M_{\text{Jennifer}} = 3.88$, $SD_{\text{Jennifer}} = 0.63$; $M_{\text{Web-search engine}} = 3.64$, $SD_{\text{Web-search engine}} = 0.82$; $\beta$ = 0.24, SE = 0.10, t = 2.47, p < 0.05* ) indicating that they were willing to trust the information provided by {\tt Jennifer} more than the information that they found from the Web-search engine and that they were more likely to use {\tt Jennifer} again. The effect size is small (Cohen's d = 0.33). 

We believe two reasons may drive people's trust. First, expert volunteers took information from credible sources, including the CDC, the WHO, and peer-reviewed journals. Second, in the conversation, {\tt Jennifer} highlights the information source and provides the link to the source in her responses. A few participants appreciated such a design. For example, \textit{``I think her sources are reliable, so I trust the chatbot. It made it easier. '' }[P65]; \textit{``...  I additionally liked how Jennifer provided the sources in her responses. ..''}[P70]. Although the Web-search engine prioritizes information from trusted sources, the result page compiles information from numerous sources which requires people to compare and identify trusted ones.

\begin{quote}
    \textit{``Completing the task via the Web-search engine required more evaluation of different sources. I had to recognize which sources were trustworthy (e.g. were they from a government institution or medical authority) and I had to scan the page to understand the context of the information presented. For more challenging questions like whether Ivermectin is a possible COVID treatment, I had to skim a medical journal article to see how they qualified their "may be effective" statement in the summary''}[P68]
\end{quote}

Second, prior studies indicated that people are willing to trust the chatbot during information exchange \cite{xiao2020tell}. The anthropomorphic features of the {\tt Jennifer} may increase the trust of the information seeker \cite{nass1994computers}. One participant commented, \textit{``The Chatbot's assistance during the quiz was very useful. Because she acted like a best friend. It replies to me like a human. I trust her''}[P56].

\begin{figure}[t]

\begin{subfigure}{0.4\textwidth}
    \includegraphics[width=0.9\linewidth]{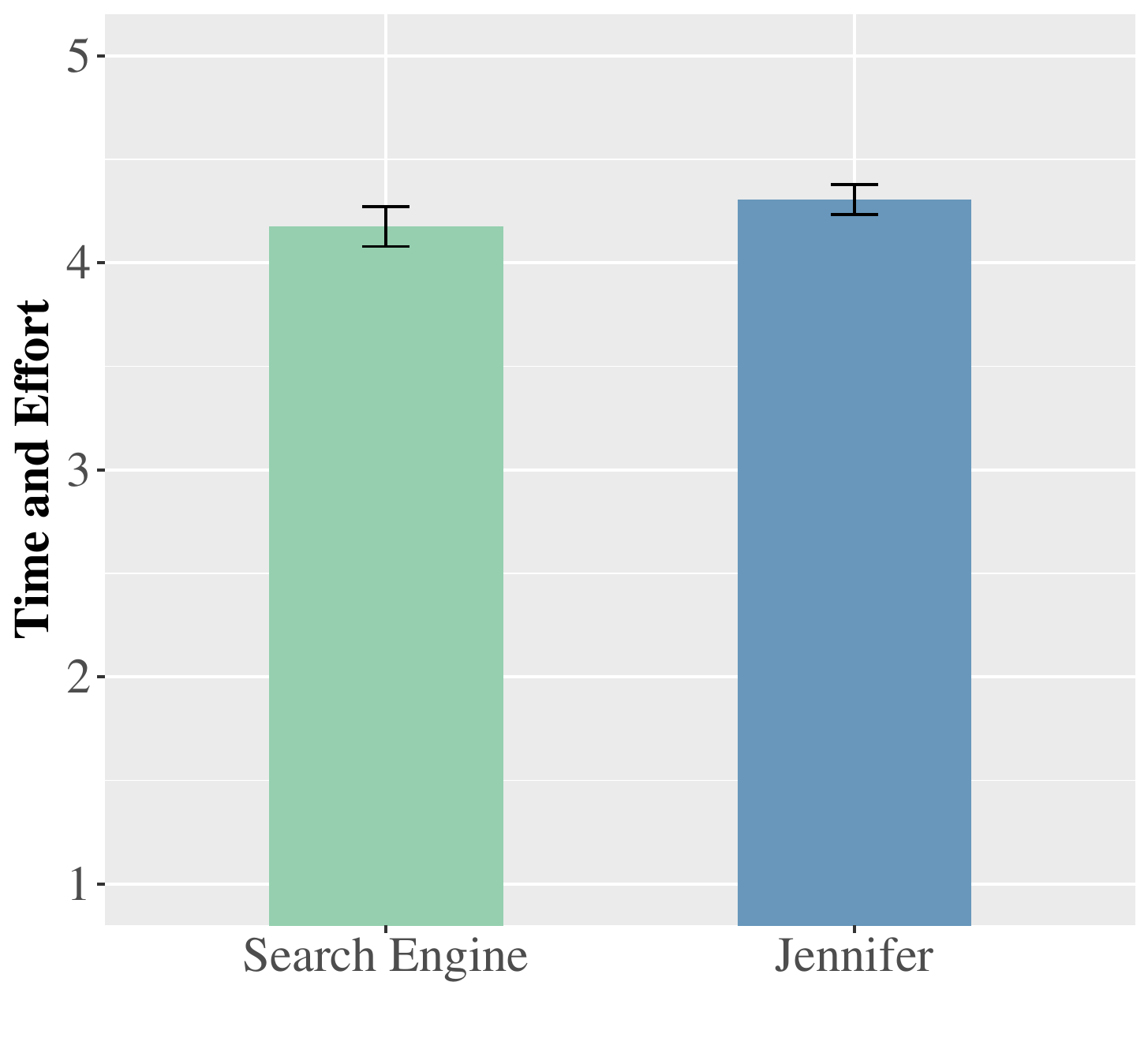}
    \caption{Perceived Time and Effort}
    \label{fig:satisfaction}
\end{subfigure}
\begin{subfigure}{0.4\textwidth}
    \includegraphics[width=0.9\linewidth]{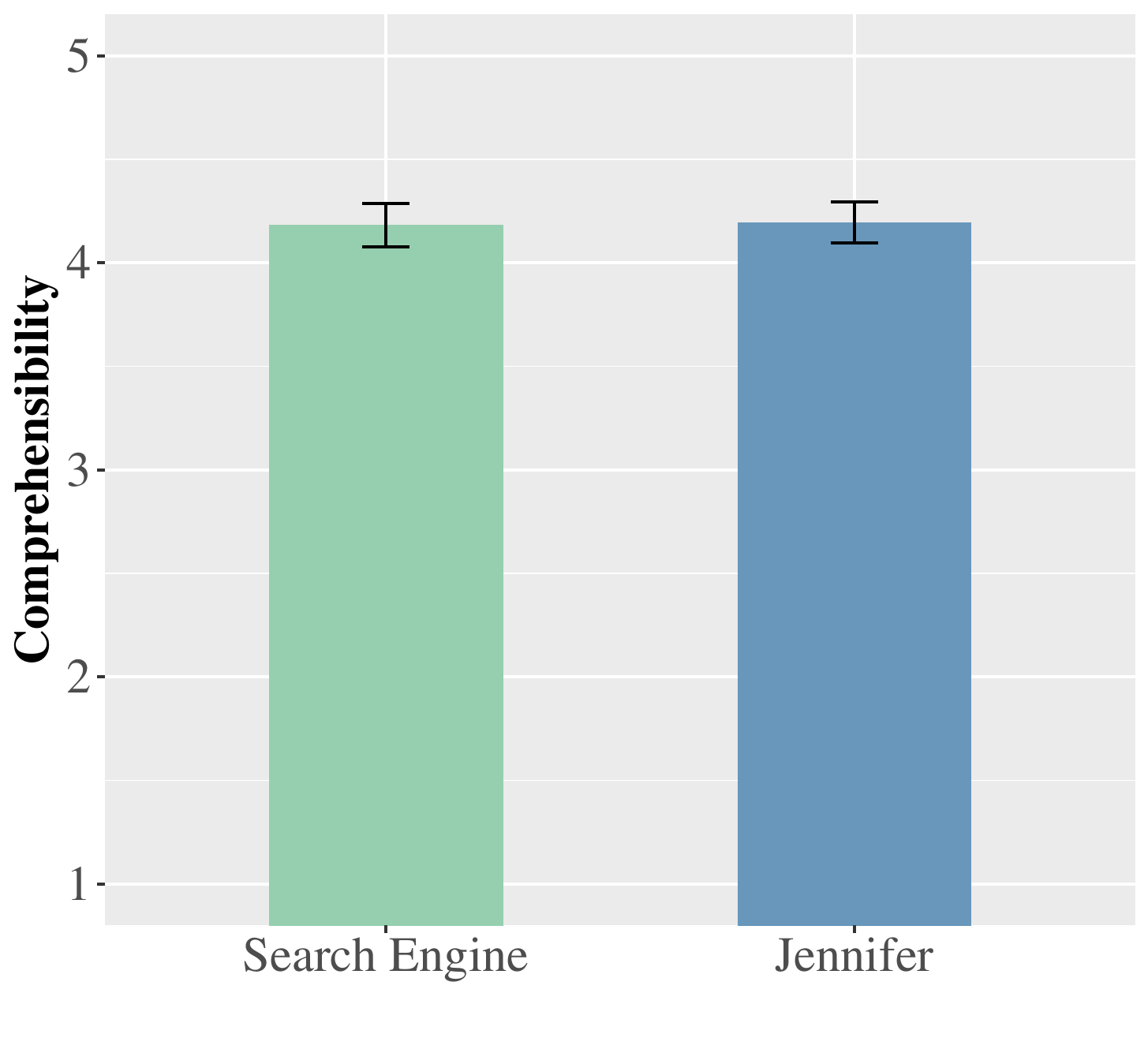}
    \label{fig:understand}
    \caption{Perceived Comprehensibility}
\end{subfigure}

\caption{The figures show contrasts between two conditions in terms of perceived time and effort and information comprehensibility. Main Findings: Our participants were satisfied with the amount of time and effort used when seeking information from {\tt Jennifer} and, compared to the Web-search engine, our participants found the information from {\tt Jennifer} was easier to understand.}
\label{fig:result2}
\end{figure}
\subsubsection{Getting information from {\tt Jennifer} is easy.}

Another dimension of effective information-seeking is time and effort. Excessive time and effort required in the process may cause information overload and disengagement. Information seekers' satisfaction also predicts future use. Our participants reported that they are overall satisfied with {\tt Jennifer} in terms of time and effort (M = 4.31; SD = 0.64). Our model shows no significant difference between Jennifer (M = 4.31; SD = 0.64) and the Web-search engine (M = 4.18; SD = 0.85; $\beta$ = 0.10, SE = 0.10, t = 1.10, p = 0.27) (Fig~\ref{fig:result2}).

\x{We believe two aspects of the {\tt Jennifer} drove satisfaction. First, the participants can read the results directly quickly after sending their questions to {\tt Jennifer} without going through a long list of results \cite{miner2020chatbots}.} A number of participants liked this quick and direct interaction, e.g., \textit{``The chatbot responded quickly to my questions. I liked how I was able to directly get answers to my questions by just typing them in.''}[P50].

Second, the natural language interface allows the information seeker to ask questions in a more natural way, \textit{``I liked that I could ... formulate questions just like I would when speaking with another person)''} [P68]. To improve the chatbot's ability to understand the information seeker's question, Helpers in {\tt Jennifer} team spent a lot of effort in training a better natural language understanding model to generate alternative questions. However, some participants also pointed out the limited natural language understanding hinders {\tt Jennifer}'s ability to provide information. \textit{``It is hard to get specific answers. The bot misunderstood exactly what I want.''} [P66]. Also, the {\tt Jennifer} could repair the conversation by providing suggestions that some participants felt useful, \textit{``When the chatbot didn't fully understand the options it listed things that were helpful in finding a suitable answer. ''} [P5].

\subsubsection{The information provided by {\tt Jennifer} is easy to understand but a bit lengthy}

Delivering health information in an intelligible way is a key goal of effective information seeking and one of our design considerations. We asked our participants if they feel the results provided by {\tt Jennifer} could be easily understood. The results showed that people feel {\tt Jennifer}'s answers to their questions are intelligible (M = 4.19; SD = 0.87). We did not observe a significant difference between {\tt Jennifer} and the Web-search engine (M = 4.18; SD = 0.93 ; $\beta$ = 0.11, SE = 0.12, t = 1.12, p  =  0.91). \textit{``I thought the answers were well thought out and explained well.''} [P12].

However, some people mentioned that answers provided by {\tt Jennifer} are lengthy, especially when formatted in the chat window. \textit{``The bot gave very lengthy and wordy answers. And the multiple chat messages rapidly came up making it more uncomfortable to read. ''} [P74]. For example, a participant may expect a question like, \textit{``Are children at risk?''}, to have a simple yes or no answer. Our response needs not only to provide an answer, although the answer may not always exist but also to let the participant know the evidence for the answer or that the answer is subject to change. {\tt Jennifer} responded as

\begin{quote}
    \textit{``Based on the current data, nobody seems to be immune from COVID-19, including children. It is true that the number of cases in children is so far lower than the number of cases in adults. We don't know why this is. The CDC provides answers to commonly asked questions about COVID-19 in children. For those interested in recent research on the subject, a study describing infections in kids in China is available [Link].''}
\end{quote}

Although people may prefer short answers, in the context of health information, we believe a lengthy answer that clearly communicates its limitation could better inform public actions.

\section{Discussion}
\x{We presented and evaluated an expert-sourcing framework to support the general public's information-seeking during public health crises. By studying two key stakeholders of our framework, the expert volunteers and the information seekers, we will discuss design implications and potential future directions for an expert-sourcing framework that can better support the expert team and create more effective chatbots as an information portal during a public health crisis.}

\subsection{Support Experts with Technologies}
\x{Through the interview with our expert volunteers, we learned about the challenges they faced in {\tt Jennifer}'s creation process. We outline a platform with four key components to better facilitate expert volunteers in the creation process. With the following technology components, the platform aims to support expert teams, especially experts without a computer science background, in creating higher-quality information portals during a public health crisis at a larger scale but with less effort.}

\x{The first component is an information management tool that can automatically track information from various sources for recent updates. To react to the rapidly changing environment quickly, Curators need to spend a significant amount of time searching and tracking content. During the interview, Curators highlighted the challenges of tracking a variety of content and voiced the need for technological support. An information management tool could automatically track information by using APIs, RSS, or web crawling. For example, \cite{shrotri2021interactive} developed a vaccine tracking tool to keep track of the vaccine development process. Once a new piece of information is added to the question base, the Curators could receive notifications once it becomes obsolete. For some contents, such as case statistics, the tool could automatically update the information with minimal human effort.}  

\x{Echoing the expert volunteer's challenge in effective communication, the second component is a writing support tool to help Curators and Helpers translate complex health information in a comprehensible manner. Communicating complex health information to the general public requires both domain expertise and writing skills. The writing support tool could provide a step-by-step guide or real-time evaluation to help Curators and Helpers better assess their answers. Or we can consider training a language generation model with high-quality answers to produce candidate output for Curators and Helpers to pick. Similar models have been built in other FAQ domains \cite{zhu2021retrieving}.}

\x{The third component is a testing and auditing tool to ensure the deployed version is error-free. Testing and auditing are crucial processes for many machine learning models in the system. When a new QA pair is added, currently, Testers need to test both new content and old content to make sure the newly added QA pair won't cause any damage, which requires tremendous human effort. A testing and auditing tool could be developed to test all existing QA pairs thoroughly with techniques such as AI-planning \cite{bozic2019chatbot}. Such a tool could also free our Tester's time to test other behaviors of the chatbot, such as repair mechanisms that need human expertise to test.}

\x{The last component is a social space for team members to chat and share. A public health crisis influences every single individual, including our expert volunteers. During the interview, many expert volunteers value social interaction among team members as a coping strategy for their emotional well-being. To support such interpersonal communications, a virtual social space not only facilitates a sense of connection within teams but also encourages global collaborations among scientists and professionals.}

\subsection{Support Information Seekers with AI Chatbots}
\x{Our expert volunteers built an AI chatbot, {\tt Jennifer}, with the proposed expert-sourcing framework to support individual seekers. In our evaluation, we found {\tt Jennifer} could effectively satisfy people's information requests, and most importantly, it could also gain people's trust. Through our user's comments and conversation log, we learned that {\tt Jennifer} might benefit from the quality content, natural language interaction, and human-like design.}

\x{People enjoyed the quality answers provided by {\tt Jennifer}. With a chatbot like {\tt Jennifer}, people could get answers directly without navigating among different sources and identifying accurate information \cite{miner2020chatbots}. In our study, people indicated {\tt Jennifer}'s answers are easy to consume, and they trust the answers when {\tt Jennifer} provides links to a credible source. Andrews et al. \cite{andrews2016keeping} showed credible sources could revitalize conversation and correct misinformation at different stages of online rumor. Our framework leverages experts' efforts to ensure {\tt Jennifer}'s a quality answer which guards users' information-seeking experience. Currently, to ensure accuracy, comprehensibility, and appropriate level of empathy, answers provided by {\tt Jennifer} is either manually curated or auto-generated with manually curated templates. While it is possible to scrape FAQs automatically from reliable resources, how to use the scraped text to generate empathetic answers with little or no training data remains an open problem \cite{DBLP:conf/aaai/LiuWLXF20}, potentially solvable via approaches similar to politeness transfer \cite{politeness}. Identifying multiple resources relevant to a question and composing answers based on them in a coherent and empathetic manner is an even more challenging problem.}

Chatbot as an information portal allows information seekers to search with natural languages \cite{radlinski2017theoretical}. A chatbot could naturally encourage information exchange to clarify ambiguous information needs and help information seekers to build complex requests \cite{rosset2020leading, tavakoli2020generating}. {\tt Jennifer} could issue clarification questions if the system can not reach the matching threshold. In our real-world deployment and online experiment, some users left comments saying such interaction helps them formulate a clearer information request, e.g., ``\textit{{\tt Jennifer} helps. Her question helps me find the right question to ask.}''[P34]. Users also issued follow-up questions to retrieve more information. The information exchange potentially helped users' search effort and contributed to {\tt Jennifer}'s success. Despite the careful chat design, people are still frustrated when {\tt Jennifer} won't be able to fully understand their questions. Since first impressions can be crucial to managing user expectations \cite{xiao2021let}, opening a conversation by stating unequivocally what the machine offers, how it operates, and what happens if or when it fails may avoid the perils of over-promising and encourage users to frame their questions with more specific keywords, and simpler sentence structures.

People trusted {\tt Jennifer}'s answers. A trustworthy information portal has real-world implications for people's prevention behavior and debunking misinformation \cite{luo2004trust}. Besides {\tt Jennifer}'s quality content, the anthropomorphic features of {\tt Jennifer} may contribute to people's trust \cite{go2019humanizing,zhou2019trusting,folstad2017chatbots,lee2020hear}. {\tt Jennifer} delivered human-like conversations that simulate social interactions, which not only delivers an engaging experience but builds rapport. The trust between the information seekers and {\tt Jennifer} makes information seekers more receptive to {\tt Jennifer}'s answer, which is especially important when they have been exposed to misinformation. In the real-world deployment, we found people ask {\tt Jennifer} for verification purposes. However, the trust may build inappropriate reliance on the information portal \cite{dzindolet2003role}, which inhibits people's ability to discern misinformation when the chatbot's answer is problematic. We should be more cautious if malicious actors could take advantage of users' trust and spread misinformation \cite{gao2018label,bursztyn2020misinformation,tran2020investigation}. We need always to keep people alert about the risk of misinformation and disinformation and avoid over-reliance on the chatbot.

\subsection{Apply the framework for the next crisis}
\x{Coordinating the distribution of information at the national level is critical to preparing for the next pandemic \cite{AlexandreWhitepaper}. Our experience with {\tt Jennifer} confirms that it is possible to collaboratively build such chatbots quickly and effectively and to scale these initiatives with the help of expert volunteers. Here, we share the lessons learned from this real-world operation of our framework and its 6-month real-world deployment.}

\textit{People are eager to help.} \x{We successfully recruited 159 experts around the globe through our own social network and newsletter. Many scientists and health professionals were eager to step up and help to better respond to the COVID-19 crisis. We found two strong motives behind their commendable efforts, altruism and the need for social support. For example, one told us, \textit{`` I always love to help out. And I want to about I don't want to be just sitting and doing nothing. ... knowing that I'm helping to create Jennifer, that was very rewarding.''}. Second, the feeling of isolation during a pandemic challenges everyone's emotional well-being. Several team members mentioned working as a team provides a sense of connection that benefits their emotional well-being. For future applications, the team should try to reach a broader community and actively support everyone's needs.}       

\textit{Effective and Dedicated Management is Critical.} \x{Even with delineation and process optimization, managing the entire process requires constant focus and dedication by a few individuals to ensure successful execution. Said one Admin, \textit{``What I have found is for pretty much the entire March and April and a great part of May, I'm spending about 20 hours every week, and just making sure everything's doing right''} [Ex5]. As such, we need to support the operation of our framework with more dedicated resources along with its large number of volunteers to ensure its long-term success.}

\textit{Process and Communication is Important.} Given the evolving tasks and a large number of volunteers with diverse backgrounds, putting the right process around tasks, workflow, and sequencing \cite{norheim2010crowdsourcing} is key to ensuring efficient use of the volunteers’ time to the advantage of the project. It is also important to hold regular dialog with the volunteers to both provide and obtain feedback as well as keep them posted about the progress of the project.

\x{\textit{Expert-in-the-loop is the Key.} Much of the recent research has focused on automating the task of fact-checking (e.g.,  ~\cite{factchecking-compjour17, pathak-srihari-2019-breaking}). However, in a novel crisis like COVID-19, facts are quickly changing. It is crucial to engage human experts in the loop to ensure the timeliness and accuracy of the answers provided by information portals like {\tt Jennifer}. Though receiving input from a large number of distributed expert volunteers is desirable, it remains an open challenge to design, construct, and maintain a fact-checking platform that supports a rigorous process to engage a large number of experts with diverse expertise levels and leverage automation in minimizing human efforts~\cite{hughes15}.}

\section{Future Work}
\subsection{Personalized Information Seeking Experience}
As the system becomes more intelligent, it opens more opportunities for personalization. A personalized system could help information seekers to refine query sentences, find more relevant information, and better consume information. In our study and real-world deployment, we found people enjoy the personalized conversation with {\tt Jennifer}. Information seekers love how {\tt Jennifer} asks clarification questions to refine their query. 

For future work, we could enable a more personalized experience by analyzing the conversation and building user representation on the fly. For example, when an information seeker asks about COVID-19 cases, the system could retrieve the case number based on their location. Or the system could also deliver medical information tailored to an information seeker's background for more effective communication. 

\subsection{AI Chatbot beyond Answering Questions}
\x{During a public health crisis, people may need more than credible information. When a crisis emerges, uncertainty is high, and there are no effective means of addressing the situation; knowing these facts may produce fear, and anxiety \cite{dillard2020fear,rogers1983cognitive}. As information seekers gather more information regarding COVID-19 through {\tt Jennifer}, especially at the early stage, it may yield higher levels of fright \cite{holman2014media}. Therefore, in the future, we should study how to provide emotional support to information seekers.}

When examining user logs, we found people were disclosing themselves and looking for emotional support while chatting with {\tt Jennifer}. For example, users told {\tt Jennifer} about challenges they are facing, \textit{``I am so sad. I lost my job''}, or emotions \textit{``I'm scared''}. This is a critical moment to provide emotional support when the user is in need. It is especially important during a crisis where many people are experiencing different kinds of hardship every day.

\x{Although the demonstration of empathy is one important design consideration, conversational skills, such as active listening skills, might be helpful to respond to users' needs and give them emotional support, or to mitigate negative psychological feelings. When necessary, it could also ask human experts to intervene \cite{lee2020designing}. For example, many chatbots have been built to support people's emotional needs, including some specifically designed for COVID-19 \cite{oh2017chatbot, martin2020artificial}.}

\subsection{Support Repeated Information Seeking with Proactive Design}
As the information becomes dynamic, the information-seeking behavior in this prolonged uncertain period is no longer single-shot. People often repeatedly seek information in reaction to the changing situation during a crisis \cite{foster2004nonlinear,palen2007crisis}. For example, Keller et al. \cite{kellar2007field} found information seekers often monitor websites to gather the most updated information. During our six-month deployment, we noticed that some users came back to {\tt Jennifer} frequently and often asked the same question for the most updated information.

In our framework, the role of {\tt Jennifer} is mostly reactive. To support such repeated information-seeking, we should consider a proactive approach. In the context of conversational agents, proactive interaction means an agent initiates interaction and drives the conversation \cite{liao2016can}. In reacting to the dynamic environment, our expert volunteers look for obsolete information and update it frequently. To further reduce information seekers' efforts, we could let information seekers opt-in for notification. Once an obsolete piece of information is updated, {\tt Jennifer} could proactively send notifications to the information seekers who asked about it in the past.

\subsection{Leverage Large Language Models}
Recent large language models (LLM) such as GPT-3 \cite{brown2020gpt3} demonstrate promising capabilities in creating engaging conversational agents, even answering people's questions \cite{ouyang2022training}. The pre-trained nature of an LLM also enables fast deployment. However, when answering people's questions, a large language model may deliver nonfactual information in a confident tone \cite{ouyang2022training}. The hallucinated answers may put information seekers at risk, especially under high-stake contexts, e.g., information regarding an ongoing public health crisis. In addition, a pre-trained LLM also has a fixed knowledge base which makes it unable to react to fast-changing situations. In future work, we ought to study how we could incorporate LLMs into our expert-sourcing framework to support information seekers with credible information while delivering engaging experiences. For example, we could use LLMs to handle people's non-health-information requests, create paraphrases of QA pairs, or regulate LLMs' nonfactual output with the expert-curated external knowledge base.
\section{Conclusions}
During a public health crisis, a credible and intelligible public health information portal could facilitate people's information-seeking efforts and inform the general public's behavior on self-protection. We presented an expert-sourcing framework to create an AI chatbot to support the general public's information-seeking in reaction to a public health crisis. We applied our framework in the real world during COVID-19, sourced global scientific communities, and created {\tt Jennifer}, an AI chatbot, to answer people's questions about COVID-19. To evaluate our framework and inform future development, we studied two key stakeholders of our framework, expert volunteers who applied our framework and built and maintained {\tt Jennifer} and information seekers who interact with {\tt Jennifer} for their information needs. Through an interview study with experts who contributed to the creation process of {\tt Jennifer}, we identified major challenges and opportunities for technology support, including a tracking system that can support content updating, a writing assistant for intelligible content, a chatbot testing environment, and a virtual space for volunteers to provide emotional support. We then conducted an online experiment to examine the effectiveness of {\tt Jennifer} in supporting information seekers' needs. The results showed that {\tt Jennifer} can effectively help information seekers retrieve the information they need, drive higher user satisfaction, and gain their trust. Our work showed an effective expert-sourcing framework to create AI chatbots as credible and easy-to-access information portals during public health crises. We further discussed our expert-sourcing framework could be applied to broader settings and future directions to support information seekers with more personalized experiences. 

\section{Acknowledgements}
We would like to thank all volunteers whose efforts have made {\tt Jennifer} possible, and the anonymous reviewers for their feedback.

\bibliographystyle{ACM-Reference-Format}
\bibliography{chatbot_health_info.bib}


\end{document}